\documentclass[12pt]{article}
\usepackage{geometry}
\usepackage{amssymb}
\usepackage{amsmath,bm}
\usepackage{amsfonts}
\usepackage{accents}
\usepackage{bbm}
\usepackage{graphicx,epsfig,lscape}
\usepackage{color}
\usepackage{setspace}
\usepackage{url}
\usepackage{upgreek}
\usepackage{commath}
\usepackage{enumitem}
\usepackage{subfigure}
\usepackage{stackengine}
\usepackage[title]{appendix}

\usepackage{indentfirst}
\usepackage{xr}
\usepackage[colorlinks=true,allcolors=blue]{hyperref}
\usepackage{natbib}
\usepackage{dsfont}
\usepackage{cleveref}
\usepackage{titlesec}
\usepackage{booktabs}
\usepackage{titling}
\usepackage{multibib}
\usepackage{multirow}
\usepackage{makecell}
\usepackage{soul}

\newtheorem{assumption}{Assumption}

\newtheorem{claim}{Claim}

\newtheorem{condition}{Condition}

\newtheorem{corollary}{Corollary}

\newtheorem{example}{Example}

\newtheorem{lemma}{Lemma}

\newtheorem{proposition}{Proposition}

\newenvironment{proof}[1][Proof]{\noindent\textbf{#1.} }{\ \rule{0.5em}{0.5em}}
\newcommand{\I}{\mathcal{I}}
\newcommand{\A}{\mathcal{A}}

\newcommand{\1}{\mathbbm1}

\newcommand{\pr}{\text{Pr}}

\newcommand\ubar[1]{\stackunder[1.5pt]{$#1$}{\rule{1ex}{.1ex}}}

\newenvironment{proprep}[1]
  {\renewcommand{\theproposition}{\ref*{#1}$'$}%
   \addtocounter{proposition}{-1}%
   \begin{proposition}}
  {\end{proposition}}

\emergencystretch=\maxdimen
\begin{document}
\author{Shanglyu Deng\thanks{
Department of Economics, University of Maryland, 3114 Tydings Hall, 7343
Preinkert Dr., College Park, MD 20742. Email: \href{mailto:sdeng1@umd.edu}{%
sdeng1@umd.edu}. I am grateful to Lawrence Ausubel for his support and guidance throughout the development of this paper. I thank Daniel Vincent, Oleg Baranov, Isa Hafalir, and Sueyoul Kim for helpful discussions, suggestions, and comments. Any errors are my own.
}}

\title{Speculation in Procurement Auctions
}
\begin{titlepage}
\maketitle

\vspace{-10pt}
\begin{abstract}
A speculator can take advantage of a procurement auction by acquiring items for sale before the auction. The accumulated market power can then be exercised in the auction and may lead to a large enough gain to cover the acquisition costs. I show that speculation always generates a positive expected profit in second-price auctions but could be unprofitable in first-price auctions. In the case where speculation is profitable in first-price auctions, it is more profitable in second-price auctions. This comparison in profitability is driven by different competition patterns in the two auction mechanisms. In terms of welfare, speculation causes private value destruction and harms efficiency. Sellers benefit from the acquisition offer made by the speculator. Therefore, speculation comes at the expense of the auctioneer.
\begin{description}
\item[Keywords:] speculation, procurement, auction theory, acquisition, supply reduction, supply withholding.

\item[JEL Classification Codes:]  C72, D44, D84, L41.
\end{description}
\end{abstract}

\thispagestyle{empty}
\end{titlepage}

\section{Introduction}
\subsection{Motivation}
In anticipation of a procurement auction, speculators have the incentive to consolidate the market by acquiring items from different sellers. By doing so, speculators gain market power and can reduce competition in the procurement auction. Consequently, a speculator's expected gain from the procurement auction may be high and more than enough to cover the acquisition costs. 

The speculative incentive offers a compelling explanation for the activities of some private equity firms around the time of the Federal Communications Commission's (FCC) Broadcast Incentive Auction. LucasPoint Networks, NRJ TV, and OTA Broadcasting acquired 48 TV stations in the pre-auction stage (\citet{doraszelski_ownership_2017}). To secure high transaction prices for the stations they actually sold in the auction, they withheld supply from the auction when the prices offered were still very high.
Shortly after the auction, the firms resold many stations they refrained from selling in the auction. The resale price of a station is often much lower than the price at which the station is dropped out, which suggests that the firms dropped out of bidding at high prices not because they value the stations more, but due to speculative reasons.

To study the profitability and welfare implications of such a speculation scheme, I incorporate a speculator and a pre-auction acquisition stage in an independent private value procurement auction model. To capture the economic insights within a parsimonious setting, I consider a single-object auction.
In the model, the speculator is allowed to make a take-it-or-leave-it offer to every seller before the auction to buy the items they have for sale.\footnote{In the case of contract procurement, the obligation to fulfill the contract is interpreted as an item for sale.}

I study speculation in both second-price auctions, which are the single-object analog to the deferred-acceptance auction design used by the FCC in the incentive auction,\footnote{To be precise, (reverse) English auctions are the single-object analog to the deferred-acceptance clock auction used by the FCC in the incentive auction. But for the purpose of this paper, English auctions and second-price auctions are equivalent.} and first-price auctions, which are adopted commonly in government procurement.

\subsection{Overview of the Results}
 The sellers weigh their prospects in the auction against the speculator's offer in deciding whether to sell to the speculator. As will be shown later, 
sellers who value their items less tend to accept the speculator's offer in equilibrium. This 
may appear to be counter-intuitive at first glance, since those sellers are more likely to win in a procurement auction than the sellers with high valuations for their items. But the comparison is misleading, as low-valuation sellers also have higher net gains than high-valuation sellers if they sell to the speculator.
Notably, in the incentive auction example, \citet{doraszelski_ownership_2017} report that ``[f]ew of the 48 TV stations [acquired by the private equity firms] are affiliated with major networks and many of them are failing or in financial distress.''

% To examine the profitability of the speculation scheme,
% it is worth noting that the speculator's profit can be decomposed into three parts: the gain from competition reduction in the auction, overcompensation for the sellers who accept the acquisition offer, 
% %% above the expected payoff they would have obtained in the auction, 
% and the loss from private value destruction when more than one items are acquired. 
% \begin{align*}
% \underline{Total Surplus} &= \underline{AuctioneerSurplus} + \underline{SellerSurplus}\\
% \underline{Total Surplus} - Efficiency Loss  &= \underline{AuctioneerSurplus} - Loss from Competition Reduction 
% \\&~~~~+ \underline{SellerSurplus} + Overcompensation
% \\&~~~~+ SpeculatorSurplus
% \end{align*}

After characterizing the equilibrium of the speculation game,\footnote{One challenge in characterizing equilibrium in the first-price auction case is to deal with asymmetric auction subgames in which a speculator bids against some seller(s). I make use of the particular form of bidder asymmetry to characterize the equilibria in the auction subgames.}
I show that the profitability of the speculation scheme hinges on the auction format. In second-price auctions, 
the speculator can always obtain a positive expected payoff by optimally choosing the acquisition price offered to the sellers. In contrast, speculation could be unprofitable in first-price auctions.
%%\footnote{The necessary and sufficient condition for speculation to be unprofitable in first-price auctions is given in the main text.}
Furthermore, the expected profit from speculation is always higher in second-price auctions than in first-price auctions. 
To understand the contrast, first note that the speculator would be a strong bidder as compared with other sellers in the auction subgame, since he has no private value for the item(s). It has been recognized in the literature that a strong bidder would often be better off in second-price auctions vis-a-vis first-price auctions (see, for example, \citet{maskin2000asymmetric}). This result, in turn, can be traced back to the different competition patterns under the two auction formats. 
%Consider an auction subgame in which the speculator bids against some sellers who rejected the acquisition offer. 
In a second-price auction, the sellers do not respond to the speculator's entry in the auction, they still bid their true values as if no acquisition had happened. However, in a first-price auction, sellers respond to the speculator's entry in the auction by bidding more aggressively. This undermines the speculator's effort in competition reduction and lessens the speculator's profit.
%\footnote{This is reminiscent of the result in \citet{salant1983losses}, which shows that horizontal merger can induce a loss when competitors' strategic responses are considered.} 
Moreover, aggressive bidding in the subgame results in a better prospect for sellers if they reject the acquisition offer and proceed to the auction. As a result, the speculator has to incur higher acquisition costs. Lower gain in competition reduction and higher acquisition costs account for the difference in speculation profitability across the two auction formats. 

The presence of a speculator has non-trivial welfare implications.
Since the speculator acquires multiple items solely for the purpose of selling one (at a high price) in the auction, the private values of all but one seller who sell to the speculator are wasted. This implies that speculation activities would induce efficiency losses. 
Sellers benefit from the presence of the speculator as the speculator overcompensates them for 
 their loss
 of private values and for skipping the auction.  This means that the speculator's profit comes at the auctioneer's expense. 
Therefore, in the case where speculation is unprofitable in first-price auctions, the auctioneer would prefer the first-price rule to the second-price rule. In this case, first-price auctions also outperform second-price auctions in terms of efficiency.

Two extensions to the second-price auction model are considered. The first one studies the profitability of the speculation scheme in a general setup where sellers' values are drawn from different distributions, and the speculator can only make acquisition offers to a subset of the sellers. Under quite general conditions, the expected profit from speculation is positive if and only if the speculator can access at least two of the sellers. This is not surprising, since the speculator can reduce competition in the procurement auction only if he has the chance to posses more than one item. 
In the second extension, I consider an enhanced speculation scheme, in which the speculator auctions off the extra items acquired back to the sellers after the procurement auction concludes.
%\footnote{To distinguish between the procurement auction and the speculator's auction, the former one is referred to as the target auction and the latter one as the ``return and refund'' auction.} 
This fine-tuned approach, if feasible,\footnote{For this approach to be feasible, either the items are homogeneous, or the timing allows the speculator to run the ``return and refund'' auction before delivering an item to the auctioneer. In the latter case, bidding in the procurement auction also needs to be for a generic item rather than a specific one. 
The feasibility of the approach is discussed in greater detail in \Cref{enhanced}.
% Moreover, the approach requires that the speculator's acquisition activities and the procurement auction happen in a relatively short time period. Otherwise, the sellers' private values may cease to exist.
} generates even more expected profit for the speculator. 

\subsection{Related Literature}
This paper is related to the literature on strategic demand/supply reduction in auctions (see, for example, \citet{vickrey1961counterspeculation}, \citet{ausubel2014demand}, and \citet{doraszelski_ownership_2017}). In the procurement context, previous studies typically take the pre-auction ownership structure of goods as given.\footnote{In the studies of multi-unit forward auctions, the demand structure of goods is typically taken as given.} This paper makes an effort to endogenize the ownership structure by considering a pre-auction acquisition stage. It is worth highlighting that in this case, efficiency can no longer be restored by a VCG auction, because it does not eliminate the speculator's incentive to accumulate market power. Although the VCG auction would superficially solve the problem of supply reduction, as the speculator would supply all the items he owns to the auction.

A growing literature is concerned with speculation in auctions with resale. In this literature, the speculator seeks to gain market power in the resale stage against other bidders, rather than in the auction stage and against the auctioneer. 
\citet{garratt2006speculation} 
%  consider the presence of a speculator in first-price or second-price auctions with resale. The speculator, who has no value for the item sold in the auction, bids to win the item in the hope of reselling it to another bidder. The authors 
show that in second-price auctions, there exist multiple equilibria in which the speculator wins the auction and makes positive profits, whereas speculation is unprofitable in first-price auctions. \citet{pagnozzi2010speculators} studies speculation in complete information multi-object auctions with resale and whether the presence of speculators helps raise the auctioneer's revenue. 
\citet{saral2012speculation}, \citet{pagnozzi2019entry}, and \citet{garratt2021auctions} experimentally study the effects of speculation driven by post-auction resale.

Also related to this paper is the bidder collusion literature, which focuses on \emph{bidders'} endeavor to reduce prospective competition in the auction. A strand of the bidder collusion literature studies 
the bidding ring's collusive arrangement while assuming that a non-strategic third-party coordinates the ring members' bidding and/or sidepayments (see, for example, \citet{graham1987collusive}, \citet{mailath1991collusion}, and \citet{mcafee1992bidding}). Another strand of this literature, forsaking the third-party approach, considers simple collusive plans proposed by a bidder. \citet{esHo2004bribing},
\citet{rachmilevitch2013bribing}, \citet{rachmilevitch2015bribing}, and \citet{troyan2017collusion} allow a bidder to make a take-it-or-leave-it bribe to his opponent in exchange for the opponent's abstain from (or bidding 0 in) the auction. \citet{lu2021perfect} allow two-option proposals which include a bribe (for the opponent to abstain) and also a request (for the proposer to abstain). Technical challenges arise as a bidder's offer may leak his private information to the opponent. In response, the authors commonly restrict their attention to second-price auctions with two collusive bidders.\footnote{One exception is \citet{rachmilevitch2013bribing}, in which two-bidder first-price auctions are considered.}$^,$\footnote{In this paper, I consider both first-price auctions and second-price auctions with $N\geq2$ bidders.
} Since I also consider the case where a strategic player (the speculator) makes simple proposals in the hope of ``gaming the system,'' the current paper is more closely related to the latter strand of the bidder collusion literature. One distinctive feature of the current paper is that the speculation scheme does not rely on (potential) bidders' commitment power, whereas the collusive scheme depends crucially on bidders' ability to commit to abstain.

The remainder of the paper is organized as follows. \Cref{model} lays out the model of a dynamic speculation game. \Cref{section spa} characterizes and analyzes the equilibrium in the second-price auction case. \Cref{section fpa} contains equilibrium characterization and analysis in the first-price auction case. \Cref{section comparison} juxtaposes the results obtained under different auction formats.  \Cref{section extensions} includes two extensions to the baseline model. \Cref{section conclusion} concludes.

\section{Model}\label{model}
An auctioneer seeks to buy an item through a reverse auction, which can be a second-price sealed bid auction (SPA), or a first-price sealed bid auction (FPA). 
The reserve price of the procurement auction is $r\in(0,1]$. 
$N\geq2$ risk-neutral sellers, indexed by the set $\I:=\{1,2,\cdots,N\}$, each has one such item for sale. Sellers' items need not be homogeneous as long as they are perfect substitutes for the auctioneer. 
Seller $i$'s value for the item is denoted by $v_i$, which is 
privately known to the seller.\footnote{Although the model is phrased for the procurement of goods and commodities, it can be applied in the case of contract procurement by interpreting the items as contractual obligations and the sellers' values as their costs.}
The sellers' values are identically and independently
distributed on $[0,1]$ according to the CDF $F(\cdot)$. The corresponding PDF is denoted by $f(\cdot)$. Throughout the paper, I assume that $F(0)=0<F(v)$ for all $v\in(0,1]$.
A seller's payoff is normalized to be 0 if no transaction happens.\footnote{One can view the private value of a seller as the opportunity cost if the seller chooses to sell the item. If no transaction occurs, the seller would not incur this opportunity cost and would get a zero payoff. On the other hand, if the item is sold at a price of $p$, the opportunity cost is incurred and the payoff to the seller is $p-v$.} 

A risk-neutral speculator, who has no item for sale and no value for any item, seeks to extract some surplus from the auction.\footnote{That the speculator has a value of 0 for an item is merely a normalization. 
What matters is that sellers' private values are surely higher than the speculator's value (either from using the good, or from reselling the good to a third-party after the auction).
This assumption seems innocuous given that the focus of the paper is on strategic rent-seeking, rather than value discovery.
}
To investigate 
the profitability and welfare implications of speculation, I consider the following two-stage game. 
At the beginning of the first stage, the speculator makes a take-it-or-leave-it offer to every seller to buy his item at a price of $p$.\footnote{Since the sellers are symmetric, I assume the speculator offers the same price to all of them. In \Cref{section extensions}, I relax the seller symmetry assumption and consider individualized acquisition offers.} Sellers choose simultaneously and individually between selling the item to the speculator, or rejecting the offer and then participating in the auction. The first stage ends after the sellers made their decisions.

In the second stage, the auction takes place. When the auction begins, the number of bidders and whether the speculator participates in the auction are announced to the bidders. Formally, the public history at the beginning of the auction is given by $h= (m,S)\in\{0,1,\cdots,N\}\times\{0,1\}$, where $m$ is the number of sellers in the auction, and $S$ describes the speculator's status, with $S=1$ standing for ``in the auction"  and $S=0$ for ``not in the auction."
If the speculator failed to buy any item in the first stage, he exits the game with zero profit. If at least one seller sold his item to the speculator, the speculator participates in the auction along with all the sellers who still possess an item. 
After the auction, if the speculator owns any surplus item(s), he uses the item(s) and derives a value of 0, or sells the item(s) to a third-party at a price of 0.\footnote{Put differently, the speculator's value for the item is determined by his best alternative after the auction, and I normalize that value to be 0.}$^,$\footnote{In the incentive auction context, if a TV station is withheld from the auction by a private equity firm, the private equity firm can continue operating the TV station, or sell the license to other firms.}
% \footnote{If the items in question are contractual obligations, there are two ways to think about the practicability of the speculation scheme. First, the service specified by the contract can be resold to a third-party. For example, if the local government seeks to auction off a milk delivery contract, the speculator may sign a number of contracts with different suppliers prior to the auction. After the auction, if the speculator owns any surplus milk delivery obligations, he supplies the milk to an outside market. Second, 
% the speculator offers to pay for a fixed period of time of the contractors. For example, say the government seeks to auction off a road pavement contract, which is to be completed in a fixed period of time. Then the speculator can make an offer to each contractor to hire the contractor for that particular time period. After the auction, if the speculator is in charge of any surplus contractors, he let them stay idled (or to work for another project) and thereby make them incur their opportunity costs.}

I characterize and analyze the perfect Bayesian equilibria (PBE) of the two-stage speculation game in the following sections.

\section{Speculation in Second-Price Procurement Auctions}\label{section spa}
% \footnote{There is no consequential difference between second-price auctions and English auctions in the current setting.}
I first study the PBE of the SPA-speculation game, holding fixed an arbitrary price offer $p\in[0,r]$.\footnote{Clearly, the speculator would never offer to pay more than $r$ for an item.}
 Once the equilibrium characterization is obtained, I proceed to investigate the speculator's choice of $p$. 
\subsection{Equilibrium Characterization}
In an auction subgame, 
bidding truthfully is a weakly dominant strategy for the sellers, but not for the speculator.  
If the speculator enters the auction with multiple items, he has an incentive to strategically reduce the supply and drive up the price. In fact, it is weakly dominant for the speculator to bid 0 for one of his items and withhold the rest from the auction. By doing so, the speculator wins the auction for sure. 
The above analysis reveals the equilibrium in weakly dominant strategies of the second-price auction subgame. I focus on this equilibrium and maintain the following assumption throughout the paper. 

\begin{assumption}[Dominant strategy equilibrium in the SPA]\label{truth}
\textup{In the SPA subgame, sellers bid truthfully and the speculator bids 0 for one of his items while withholding the rest from the auction.}
\end{assumption}

\begin{lemma}\label{spa_cutoff}
Under \Cref{truth} and holding fixed $p\in[0,r]$, in any PBE of the SPA-speculation game, there exists $v^*\in[0,r]$ such that seller $i\in\I$ accepts the speculator's offer if and only if $v_i< v^*$. 
\end{lemma}

\Cref{spa_cutoff} indicates that sellers with low values tend to accept the speculator's offer in equilibrium. Notably, 
those sellers would have had good chances to win in the procurement auction but for the speculator's presence.

With \Cref{spa_cutoff}, it only remains to pin down the cutoff $v^*$ for equilibrium characterization. 
Intuitively, 
if $p$ is too low, no seller would accept the offer. 
In fact, $p$ must exceed 
\[
\pi_0:=\int_0^r[1-F(x)]^{N-1}dx,\footnote{This is the interim expected payoff of a seller with a realized value of 0 in the absence of the speculator.}
\]
to induce a nonzero probability of sellers' accepting the offer, or equivalently, to induce a positive cutoff $v^*$. If $r\geq p>\pi_0$, the following indifference condition determines the cutoff $v^*$,
\begin{equation}\label{spa_eq1}
p-v^* = \int_{v^*}^r[1-F(x)]^{N-1}dx,
\end{equation}
where the left hand side is the marginal seller's payoff from accepting the acquisition offer, and the right hand side is the payoff from participating in the auction.

The PBE is described by \Cref{spa_eqm}, with $v^*(p)$ defined as follows,
% \begin{align}
%     p^*(v)&:=v+\int_v^r[1-F(x)]^{N-1}dx,\text{ for }v\in[0,r],\text{ and }\\
% v^*(p)&:=\begin{cases}
% 0,&\text{ if }p\leq \pi_0,\\
% \text{the unique solution to }p = p^*(v),&\text{ if }\pi_0<p\leq r.
% \end{cases}
% \end{align}
% \[
% p^*(v):=v+\int_v^r[1-F(x)]^{N-1}dx,\text{ for }v\in[0,r],
% \]
% and
\begin{equation}
    v^*(p):=\begin{cases}
0,&\text{ if }p\leq \pi_0,\\
\text{the unique solution to }\eqref{spa_eq1},&\text{ if }\pi_0<p\leq r.
\end{cases}
\end{equation}
\begin{proposition}\label{spa_eqm}
Under \Cref{truth} and holding fixed $p\in[0,r]$, there exists a unique PBE of the SPA-speculation game. In the equilibrium, seller $i\in\I$ accepts the speculator's offer if and only if $v_i<v^*(p)$.
\end{proposition}
% It is worth pointing out that the PBE does not rely on the availability of public information $h\equiv(m,S)$ at the beginning of the auction. Sellers would have a strict incentive to bid seriously if 

\subsection{Analysis of the Equilibrium}
\Cref{spa_eqm} paves the way for investigating the speculator's choice of $p\in[0,r]$. Since 
there is a one-to-one mapping between the price $p$ and the cutoff $v^*$ (except for the $v^*=0$ case), 
it is more convenient to think of 
the speculator's problem as choosing a profit-maximizing cutoff level $v^*$ from $[0,r]$. 

For a fixed equilibrium cutoff $v^*\in(0,r]$, the corresponding acquisition price is 
\begin{equation}\label{spa price}
p^*(v^*):=v^*+\int_{v^*}^r[1-F(x)]^{N-1}dx.
\end{equation}
The speculator's expected profit is 
\begin{equation}\label{spa pi}
\Pi^*(v^*):= \sum_{m=0}^{N-1}{N\choose m}[F(v^*)]^{N-m}[1-F(v^*)]^my(m,v^*)-NF(v^*)p^*(v^*),
\end{equation}
with the expected payment from an auction subgame against $m$ sellers given by 
\[
y(m,v^*):=v^*+\int_{v^*}^r\left[\frac{1-F(x)}{1-F(v^*)}\right]^mdx.
\]

The speculator's expected profit consists of three parts.
First, the speculator gains from supply withholding:\footnote{To distinguish from the term ``supply reduction,'' which is typically used in the context of multi-unit auctions, I refer to the speculator's overbidding in the single-unit procurement auction as ``supply withholding.'' Although there is no substantial difference in the meaning of the two terms.} when the speculator enters the auction with multiple items, he reduces the competition in the auction by withholding all but one of his items from the auction. Second, the speculator loses from overcompensating the sellers: when the speculator purchases an item from a seller, he pays more than the expected payment that the seller would have received in an auction without speculation.\footnote{Consider a seller with realized value $v<v^*$. The expected payment the seller would have received in the absence of the speculator is $v+\int_v^r[1-F(x)]^{N-1}dx$, which is less than $p^*(v^*)$, the price paid by the speculator.} Third, the speculator loses from destroying private values: he has no use for the items he owns after the auction, but to purchase them before the auction, he must compensate the sellers for their loss of private values. 
This decomposition is most clearly seen in the two-seller case. Fix $N=2$, the speculator's expected profit is 
\[
\underbrace{[F(v^*)]^2\left\{r-\int_0^{v^*}xd\left[\frac{F(x)}{F(v^*)}\right]^2\right\}}_{\text{gain from supply withholding}}-\underbrace{2\int_0^{v^*}[F(x)]^2dx}_{\text{loss from overcompensating the sellers}}-\underbrace{\int_0^{v^*}xd[F(x)]^2}_{\text{loss from destroying private values}}.
\]
To see that the first term represents the gain from supply withholding, note that $[F(v^*)]^2$ is the probability that two sellers both accept the acquisition offer, which is the premise of supply withholding; $r$ is the payment received by the speculator with supply withholding, and; $\int_0^{v^*}xd\left[\frac{F(x)}{F(v^*)}\right]^2$ is the payment would have been made by the auctioneer in the absence of the speculator. The second term is the increment in the sellers' expected payoff in the speculator's presence, which comes from the speculator's overcompensation. 
The third term is the expected efficiency loss due to private value destruction. This loss is suffered by a seller, but is passed through to the speculator.

Notably, in the $N=2$ case, as $v^*$ approaches 0, the gain from supply withholding dominates the losses.\footnote{To be precise, the gain is of the magnitude $[F(v^*)]^2r$, whereas the losses are of the magnitude $[F(v^*)]^2v^*$, which is a higher order infinitesimal.} This implies that the speculator can always get a positive expected profit by choosing $v^*$ appropriately. \Cref{spa_profitable} generalizes this result to the $N$-seller case.   

\begin{proposition}\label{spa_profitable}
The speculator can always get a positive expected profit in the SPA-speculation game. 
\end{proposition}
In fact, as $v^*$ approaches 0, the major component of the speculator's profit is the gain from supply withholding when exactly two sellers accept the acquisition offer. Formally, 
\[
\lim_{v^*\to0}\frac{\Pi^*(v^*)}{[F(v^*)]^2} = {N\choose 2}\int_0^r[1-F(x)]^{N-2}dx.
\]
To see that the right hand side is the gain from supply withholding when exactly two sellers accept the acquisition offer, note that $\int_0^r[1-F(x)]^{N-2}dx$ is the limit of expected payment received by the speculator after supply withholding.
The payment would have been made by the auctioneer in the absence of the speculator 
does not appear 
on the right hand side
because it
is less than $v^*$, which approaches 0.

\Cref{spa_profitable} implies that the speculator would always induce a positive cutoff in the speculation game. By inspecting the equilibrium outcome, the welfare implications of speculation are immediately obtained.
\begin{corollary}\label{spa_wel}
Speculation results in an efficiency loss in the form of private value destruction. Sellers are better-off in the presence of the speculator, while the auctioneer is worse-off.
\end{corollary}

It is worth highlighting that the inefficiency is not caused by strategic supply withholding alone. If, instead of a second-price auction, a VCG auction is used for the procurement, the speculator would bid truthfully for all of his items.\footnote{For instance, if the speculator had acquired three items, he would bid $(0,0,0)$ in the VCG auction. In contrast, he effectively bids $(0,\infty,\infty)$ in the second-price auction.} There would be no strategic supply withholding in the auction, yet the equilibrium outcome of the speculation game would not change. Although VCG auctions 
can restore efficiency by eliminating strategic supply/demand reduction in a setting where the ownership structure is fixed, they cannot eliminate the incentive to become a multi-unit owner in a setting where endogenous changes to the ownership structure can happen. 
% In such a setting, inefficiency is generally unavoidable without another resale opportunity for the speculator to refund some extra items.

\section{Speculation in First-Price Procurement Auctions}\label{section fpa}
In this section, I consider speculation in first-price auctions. For simplicity, 
I restrict attention to symmetric PBE in the FPA-speculation game.
That is, sellers are assumed to use the same strategy in equilibrium. As in the previous section, I first characterize the equilibrium for any fixed $p\in[0,r]$. Then I proceed to analyze the speculator's choice of $p$. 

\subsection{Equilibrium Characterization}

\Cref{fpa cutoff} establishes the cutoff structure in equilibrium.
\begin{lemma}\label{fpa cutoff}
Holding fixed $p\in[0,r]$, in any symmetric PBE of the FPA-speculation game, there exists $v^\star\in[0,r]$ such that seller $i\in\I$ accepts the speculator's offer if and only if $v_i<v^\star$.\footnote{More precisely, there exists a $F$-measure 0 set $\mathcal E$ such that seller $i$ accepts the speculator's offer if and only if $v_i\in[0,v^\star)\setminus\mathcal E$.}
\end{lemma}

In light of \Cref{fpa cutoff},  it is useful to consider the subgame in which the speculator and $m\geq1$ sellers compete with each other. The speculator has no value for his item,\footnote{The speculator may own many items, but he will withhold all but one from the auction. } while each seller's value is independently drawn from $[v^\star,1]$, with $v^\star\in[0, r]$, according to the CDF $G(\cdot;v^\star):=\frac{F(\cdot)-F(v^\star)}{1-F(v^\star)}$. The auction subgame  is labeled as $\Gamma:=\left<r, m, G(\cdot;v^\star)\right>$. 

I restrict attention to equilibrium in undominated strategies and maintain \Cref{ubfpa} throughout the rest of the paper. 
\begin{assumption}[Undominated bidding in the FPA]\label{ubfpa}
\textup{In the FPA subgame, bidders never bid below their values.}
\end{assumption}

\Cref{fpa payoff} reveals information on the speculator's and the sellers' equilibrium payoffs in the first-price auction subgame. This result will help set up the sellers' indifference condition and characterize the PBE of the speculation game. 

\begin{lemma}\label{fpa payoff}
Under \Cref{ubfpa}, the following statements hold 
 in any symmetric BNE of the FPA subgame $\Gamma=\left<r, m, G(\cdot;v^\star)\right>$. 
\begin{enumerate}[label = (\roman*)]
\item The equilibrium payoff of the speculator (in the subgame $\Gamma$) is 
\[
\ubar b(m,v^\star):=\max_{r\geq b\geq v^\star}b[1-G(b;v^\star)]^m.\]
\item The interim equilibrium payoff of a seller with value $v^\star$ is 
$\ubar b(m,v^\star)-v^\star$.
\end{enumerate}
\end{lemma}

\Cref{fpa payoff} is obtained on the condition that one such equilibrium exists.
The issue of equilibrium existence is addressed by 
 \Cref{fpa sub}, which provides the characterization of a symmetric BNE. 

\begin{lemma}\label{fpa sub}
A symmetric BNE (in which the bidders never bid below their valuations) of the FPA subgame $\Gamma=\left<r, m, G(\cdot;v^\star)\right>$ is given as follows.  
\begin{enumerate}[label = (\roman*)]
\item The sellers bid according to 
\[
\beta(v;m,v^\star) = \begin{cases}
\ubar b(m,v^\star)/[1-G(v;v^\star)]^m,&\text{ if }v^\star\leq v\leq \bar b(m,v^\star),\\
v,&\text{ if }v>\bar b(m,v^\star),
\end{cases}
\]
where
\[\bar b(m,v^\star) := \min\left\{\text{arg}\max_{r\geq b\geq v^\star} b[1-G(b;v^\star)]^m\right\}.\]
\item The speculator mixes over $[\ubar b(m,v^\star),\bar b(m,v^\star)]$ 
 according to the CDF 
 \[
 \Psi(b) = 1-\exp\left\{-\int_{v^\star}^{\beta^{-1}(b;m,v^\star)}\frac{[\beta(x;m,v^\star)+(m-1)x]}{[\beta(x;m,v^\star)-x][1-G(x;v^\star)]}dG(x;v^\star)\right\}.
 \]
\end{enumerate}
\end{lemma}

For the marginal seller to be indifferent between accepting the acquisition offer and participating in the procurement auction, the following must hold,
	\begin{equation}\label{fpa indiff}
	p-v^\star= \int_{v^\star}^r[1-F(x)]^{N-1}dx+\sum_{m=0}^{N-2}\binom{N-1}{m}[1-F(v^\star)]^m[F(v^\star)]^{N-1-m}[\ubar b(m+1,v^\star)-v^\star].
	\end{equation}
Clearly, the left hand side is the seller's payoff from accepting the speculator's offer. The right hand side is the seller's expected payoff from the auction subgames, as indicated by \Cref{fpa payoff}.

 A one-to-one mapping between $p\in[\pi_0,r]$ and $v^\star\in[0,r]$ is defined implicitly by the indifference condition \eqref{fpa indiff}. I denote the mapping from $v^\star\in[0,r]$ to $p\in[\pi_0,r]$ by $p^\star(v^\star)$. Specifically, 
 \begin{equation}\label{fpa price}
 \begin{split}
p^\star(v^\star):=v^\star&+ \int_{v^\star}^r[1-F(x)]^{N-1}dx\\
&+\sum_{m=0}^{N-2}\binom{N-1}{m}[1-F(v^\star)]^m[F(v^\star)]^{N-1-m}[\ubar b(m+1,v^\star)-v^\star].\footnotemark
\end{split}
\end{equation}
\footnotetext{It can be verified that $p^\star(v^\star)$ is strictly increasing in $v^\star\in[0,r]$, with $p^\star(0) = \pi_0$ and $p^\star(r) = r$. As a result, the one-to-one mapping between $p\in[\pi_0,r]$ and $v^\star\in[0,r]$ is well-defined. See Appendix \ref{proofs} for details. \label{fpa pv}}

The indifference condition \eqref{fpa indiff} suggests that if $p\leq\pi_0$, no seller will accept the acquisition offer in equilibrium. I include this case in the definition of $v^\star(p)$, the inverse of $p^\star(v^\star)$. Specifically, 
\[
v^\star(p):=\begin{cases}
0,&\text{ if }p\leq \pi_0,\\
\text{the unique solution for $v^\star$ to }\eqref{fpa indiff},&\text{ if }\pi_0<p\leq r.\\
\end{cases}
\]
\Cref{fpa eqm} provides an equilibrium characterization for the speculation game. 
\begin{proposition}\label{fpa eqm}
Under \Cref{ubfpa} and holding fixed $p\in[0,r]$, the following statements hold in the FPA-speculation game.
\begin{enumerate}[label = (\roman*)]
\item  In any symmetric PBE, 
seller $i\in\I$ accepts the speculator's offer if and only if $v_i<v^\star(p)$. 
\item The cutoff acceptance strategy and the bidding strategies in \Cref{fpa sub} (with $v^\star$ replaced by $v^\star(p)$) constitute a symmetric PBE.\footnote{If all the sellers sold their items to the speculator, the speculator bids $r$ in the procurement auction; if all the sellers rejected the acquisition offer, they bid as in a standard FPA model with symmetric bidders.} 
\end{enumerate}
\end{proposition}

\subsection{Analysis of the Equilibrium}
Again, thanks to the one-to-one mapping between $p\in(\pi_0,r]$ and $v^\star\in(0,r]$, the speculator's problem of choosing $p$ can be reformulated as choosing $v^\star$. 
By \Cref{fpa payoff} and \Cref{fpa eqm}, in any symmetric PBE of the speculation game, for a given cutoff level $v^\star\in[0,r]$, the speculator's expected profit is given by 
\begin{equation}\label{fpa pi}
\Pi^\star(v^\star) := \sum_{m=1}^{N-1}\binom{N}{m}[1-F(v^\star)]^m[F(v^\star)]^{N-m}\ubar b(m,v^\star)+[F(v^\star)]^{N}r-N[F(v^\star)]p^\star(v^\star).
\end{equation}

The following result on the profitability of speculation is readily available. 
\begin{proposition}\label{fpa unprof}
Speculation could be unprofitable in first-price auctions. 
\end{proposition}
The proof of \Cref{fpa unprof}, as well as a comparative statics analysis on the profitability of speculation, is given by \Cref{fpa num}.

\begin{figure}[th!]
\begin{center}
\includegraphics[width=2.7in]{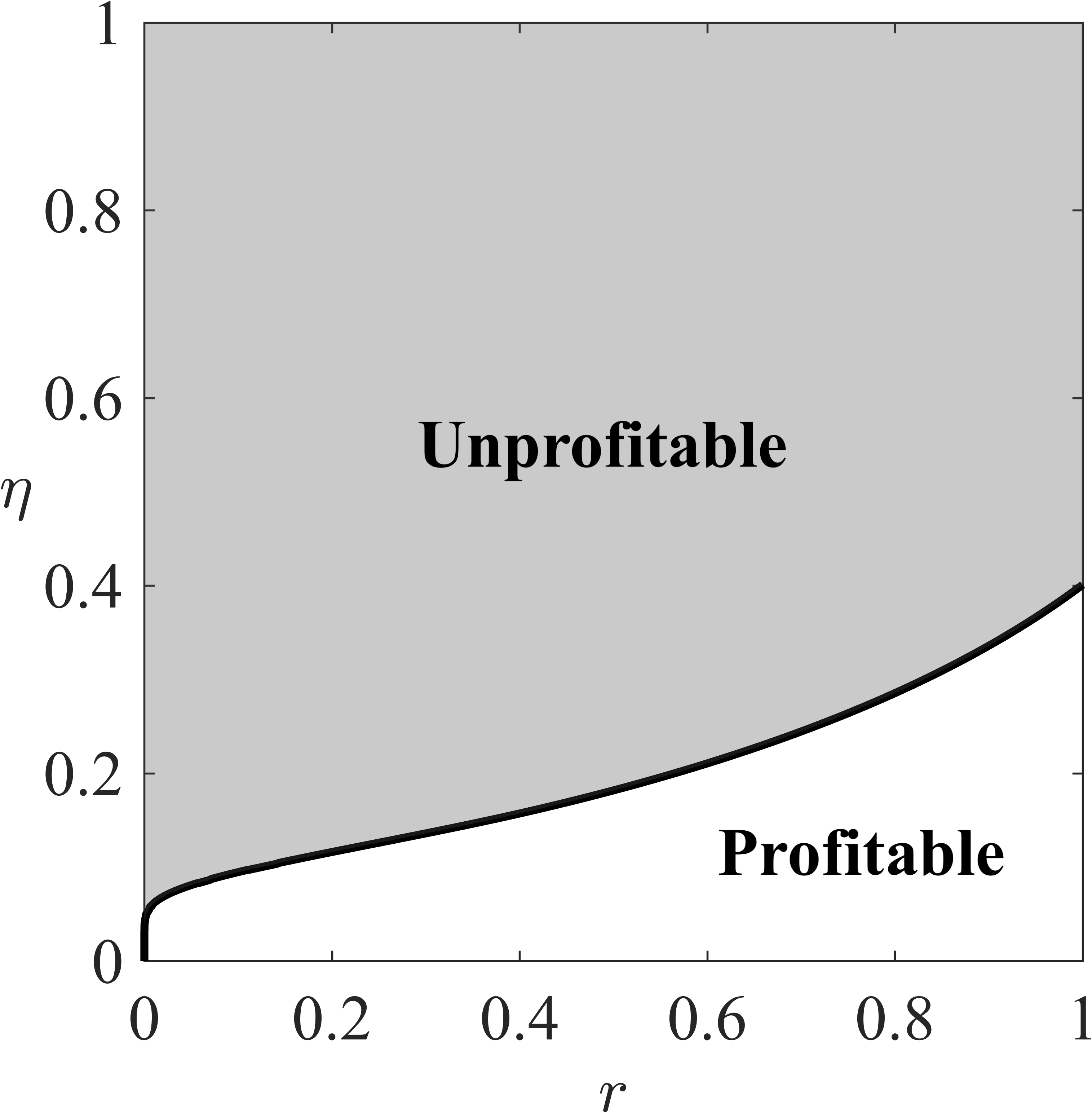}
\caption{Profitability of speculation in FPAs, with $N=2$ and $F(v) = v^\eta$.}
\label{fpa_fig}
\end{center}
\end{figure}

\begin{example}\label{fpa num}
\textup{Fix $N=2$ and let $F(v) = v^\eta$, with $\eta>0$.
Since sellers are more likely to have high valuations (and be \emph{less} competitive in the procurement auction) with a larger $\eta$, $\eta$ can be thought of as a measure of sellers' competitiveness. 
 \Cref{fpa_fig} shows how the profitability of speculation in first-price auctions changes with the reserve price $r$, and the competitiveness parameter $\eta$. }
%%%% Speculation is profitable if the $(\eta,r)$ pair is in the lower right part of figure, implying that the reserve price is high and sellers are competitive. If the $(\eta,r)$ pair is in the shaded area, speculation is unprofitable. 
\end{example}

For a fixed $\eta$, speculation is profitable if $r$ is high enough and unprofitable otherwise. 
Since the speculator's profit comes from the auctioneer, it is not surprising that a higher willingness-to-pay of the auctioneer leaves more room for speculation. 

For a fixed $r$, speculation is profitable if $\eta$ is small enough and unprofitable otherwise. 
A small $\eta$ implies that a seller is more likely to have a low valuation, so the seller would be willing to accept the acquisition offer at a low price. 
In the meantime, a small $\eta$ means that other sellers are likely to have low valuations, so the competition in the procurement auction would be fierce. This further reduces the seller's willingness-to-accept for the acquisition offer. 
As a result, the speculator benefits from sellers' competitiveness.

\section{A Comparison of the Two Auction Formats}\label{section comparison}
In this section, I compare speculation under the two auction formats and identify the driving force underlying the difference.  
\subsection{Profitability Ranking}
In the previous sections, it is shown that speculation tends to be successful in second-price auctions but could be unprofitable in first-price auctions. \Cref{more} complements the result by quantitatively comparing the expected profit from speculation  under the two auction formats.
\begin{proposition}\label{more}
Speculation is strictly more profitable in second-price auctions than in first-price auctions.
\end{proposition}
The proof of \Cref{more} is kept here in the main text since it is informative about the underpinning of the result. 

\begin{proof}
Consider the same equilibrium cutoff $v^\dagger\in(0,r)$ under the two auction formats. The speculator's expected profit is higher under the second-price rule than under the first-price rule---i.e., $\Pi^*(v^\dagger)>\Pi^\star(v^\dagger)$---for the following two reasons. 

First, the corresponding acquisition price is higher in the first-price auction case, since sellers have better prospects in the first-price auction subgames. In particular, in an auction subgame with the speculator's participation, a seller never wins under the second-price rule but could win under the first-price rule. Formally,
\[
p^\star(v^\dagger)-p^*(v^\dagger) = \sum_{m=0}^{N-2}\binom{N-1}{m}[1-F(v^\dagger)]^m[F(v^\dagger)]^{N-1-m}[\ubar b(m+1,v^\dagger)-v^\dagger],
\]
where the right hand side represents the extra expected payoff a type-$v^\dagger$ seller can get in first-price auction subgames. 

Second, the speculator is better-paid in a second-price auction subgame than in a first-price auction subgame. If the speculator is competing against $1\leq m\leq N-1$ sellers in an auction subgame summarized by
$\Gamma=\left<r, m, G(\cdot;v^\dagger)\right>$, the speculator's expected payoff is $y(m,v^\dagger)$ under the second-price rule and is $\ubar b(m,v^\dagger)$ under the first-price rule. \Cref{auction_payment} in Appendix \ref{proofs} shows that 
$
y(m,v^\dagger)>\ubar b(m,v^\dagger)\text{ for }v^\dagger\in(0,r).
$

As a result, it is easy to see that $\Pi^*(v^\dagger)>\Pi^\star(v^\dagger)$ for all $v^\dagger\in(0,r)$ by comparing equations \eqref{spa pi} and \eqref{fpa pi}. 
%%%\[
%%%\begin{split}
%%%\Pi^*(v^\dagger)-\Pi^\star(v^\dagger) =  \sum_{m=1}^{N-1}\binom{N}{m}[1-F(v^\dagger)]^m[F(v^\dagger)]^{N-m}[y(m,v^\dagger)-\ubar b(m,v^\dagger)]&\\
%%%+N[F(v^\dagger)][p^\star(v^\dagger)-p^*(v^\dagger)]>0.&
%%%\end{split}
%%%\]
An immediate implication is  
$
\max_{v^\dagger\in[0,r]}\Pi^*(v^\dagger) \geq \max_{v^\dagger\in[0,r]}\Pi^\star(v^\dagger).
$
If $\Pi^\star(v^\dagger)$ is not maximized by $0$ or $r$, the inequality is strict. If $\Pi^\star(v^\dagger)$ is maximized by $0$, the right hand side of the inequality is 0. By  \Cref{spa_profitable}, the left hand side is positive, so the inequality is strict. Finally, $\Pi^\star(v^\dagger)$ cannot be maximized by $r$, since $\Pi^\star(r) = \left\{1-[1-F(r)]^N-NF(r)\right\}r<0$. 

In summary, $\max_{v^\dagger\in[0,r]}\Pi^*(v^\dagger) > \max_{v^\dagger\in[0,r]}\Pi^\star(v^\dagger)$. This concludes the proof. 
\end{proof}

The proof of \Cref{more} builds on the difference in the auction outcomes under the two formats, which can be traced back to the difference in competition patterns. Specifically, in second-price auctions, sellers do not respond to the presence of a strong competitor, namely, the speculator: they still bid truthfully in equilibrium. However, in first-price auctions, as will be shown below, sellers respond to the presence of a strong competitor by bidding more aggressively. This helps the sellers win the procurement auction sometimes, even if their values are surely above the speculator's. Therefore, sellers have better prospects in first-price auctions, whereas the speculator's expected payoff is lower.

To see how sellers respond to the speculator's presence in first-price auction subgames, first consider the following hypothetical benchmark. Suppose that in any first-price auction subgame $\Gamma=\left<r, m, G(\cdot;v^\star)\right>$ with $1\leq m\leq N-1$, sellers bid as if the speculator was just another seller. That is, the speculator also drew his value from $[v^\star,1]$ according to $G(\cdot;v^\star)$. In this case, sellers' bidding strategy is given by 
\[
\tilde\beta(v;m,v^\star) =
\begin{cases}
v+\int_v^r[1-G(x;v^\star)]^mdx\big/[1-G(v;v^\star)]^m,&\text{ if }v^\star\leq v\leq r,
\\
v,&\text{ if }v> r.
\end{cases}
\]

This is a suitable benchmark, since it hypothetically achieves  revenue equivalence across the two auction formats. 
Specifically, the allocative outcome, the speculator's expected profit, the auctioneer's expected surplus and the sellers' expected payoffs would be the same under the two formats.\footnote{Notice that the speculator would bid $\tilde\beta(v^\star;m,v^\star)$ and always win the procurement auction in the hypothetical benchmark.}

In contrast with the benchmark, sellers bid against a stronger competitor in reality. They respond by bidding more aggressively as indicated by \Cref{aggressive}.
\begin{lemma}\label{aggressive}
In a FPA subgame, sellers bid more aggressively when they compete against the speculator than when they compete against another seller. Formally, $\beta(v;m,v^\star)<\tilde\beta(v;m,v^\star)$ for all $v\in[v^\star,r)$.
\end{lemma}
As is shown in the previous analysis, sellers' aggressive bidding undermines the profitability of speculation in first-price auctions. Difference in sellers' responses accounts for the result in \Cref{more} and also the contrast between \Cref{spa_profitable} and \Cref{fpa unprof}.

\subsection{Welfare Comparison}
With speculation, the revenue equivalence between first-price auctions and second-price auctions breaks down. 
In this case, welfare raking is generally ambiguous. 
In particular, 
although speculation is more profitable in second-price auctions, 
it does not necessarily lead to more efficiency loss. 
First-price auctions could be less efficient, as the profit-maximizing cutoff level could be higher therein. Even with the same cutoff level, a first-price auction is less efficient than a second-price auction, since it includes an additional source of inefficiency other than the destruction of private values: the allocative inefficiency in asymmetric auction subgames. 

Since an explicit expression for the profit-maximizing cutoff level is unavailable under the two formats, a general welfare ranking result cannot be obtained.  Nevertheless,
when speculation is not profitable in first-price auctions, the welfare ranking is  given by the following corollary.
\begin{corollary}
In the presence of a speculator,
provided that $\Pi^\star(v^\star)<0$ for all $v^\star\in(0,r]$, 
FPA is better for the auctioneer and efficiency but worse for the sellers than SPA.
\end{corollary}

\section{Extensions to the Second-Price Auction Model}\label{section extensions}
Two extensions to the second-price auction model are presented in this section. First, I consider a more general setting and reexamine the profitability of speculation. Specifically, 
 I relax the assumption that sellers draw their values from the same distribution. In the meantime, I consider the case where the speculator can only make offers to a subset of the sellers.
This limited access assumption is relevant when the speculator wishes to keep speculation discreet by operating on a limited scale, or when the speculator is subject to a budget constraint. 

Formally, the baseline model is modified as follows for this section.
For all $i\in\I$, seller $i$'s value $v_i$ is drawn from [0,1] according to the CDF $F_i(\cdot)$. The corresponding PDF is $f_i(\cdot)$. Again, I assume $F_i(0) = 0<F_i(v)$ for all $v\in(0,1]$. 
The speculator can make (potentially different) acquisition offers only to a subset of the sellers, denoted by $\A$. 

Second, I study an enhanced speculation approach which is fine-tuned to address the loss from private value destruction. While I focus mainly on the second-price auction case, a result regarding enhanced speculation in first-price auctions is included to facilitate comparison. 

\subsection{Speculation with Limited Access to Asymmetric Sellers}
In this part, I reexamine the profitability of speculation in second-price auctions. 

In light of the analysis of the baseline case, I consider price offers that will lead to a vector of acceptance/rejection cutoffs $\bm{v}^*:=(v_j^*)_{j\in\A}$, with $v_j^*\in[0,r]$.\footnote{Clearly, $v_j^*>r$ (which implies that $p_j>r$) is suboptimal for the speculator. }  That is, seller $j\in\A$ accepts the speculator's offer if and only if $v_j<v_j^*$. Define 
\begin{equation}\label{laas p}
p_j^*(\bm v^*):=v_j^*+\int_{v_j^*}^r\prod_{i\in\I\setminus\A}[1-F_i(x)]\prod_{s\in\A\setminus\{j\}}\left[1-F(\max\{x,v_s^*\})\right]dx.
\end{equation}
\Cref{spa eqm 2} gives the equilibrium characterization in this case.

\begin{proposition}\label{spa eqm 2}
Suppose that the speculator offers $p_j = p_j^*(\bm v^*)$ to seller $j\in\A$, with $v_j^*\in[0,r]$. Then a PBE of the SPA-speculation game is described as follows. Seller $j\in\A$ accepts the speculator's offer if and only if $v_j<v_j^*$. 
In the procurement auction, sellers bid truthfully and the speculator engages in strategic supply withholding.
\end{proposition}

Note that the speculator's profit originates from his ability to reduce potential competition in the auction. If the speculator cannot reduce the competition in a meaningful way, either because the speculator can make acquisition offer to only one seller, or because some sellers outside the speculator's reach are very competitive, speculation would not be profitable. \Cref{con1} addresses the two possibilities. 
\begin{condition}\label{con1}
There exists $\{k,k'\}\subseteq\A$, such that for all $i\in\I\setminus\A$, 
\[
\lim_{v\to0}[vF_i(v)/F_k(v)] = 0\text{ and }\lim_{v\to0}[vF_i(v)/F_{k'}(v)] = 0.
\]
\end{condition}
It is clear that \Cref{con1} rules out the case that $\A$ is a singleton. \Cref{con1} also requires 
the sellers outside the speculator's reach are not too competitive as compared with some sellers in the speculator's reach. Specifically, \Cref{con1} ensures 
the values of the sellers in $\I\setminus\A$ do not concentrate at the lower end much more than the values of some sellers in $\A$ do. This requirement does not appear to be too strong, and it is trivially met if sellers are symmetric.

\Cref{spa_profitable} can be generalized to the current setting as follows. 
\begin{proprep}{spa_profitable}
\label{asym profit}
If \Cref{con1} is satisfied, speculation in the second-price procurement auction is profitable.
\end{proprep}

\subsection{Enhanced Speculation}\label{enhanced}
In this part, I consider an enhanced speculation scheme fined-tuned to address the issue of private value destruction. To that end, I assume that after the procurement auction, the speculator can auction off his surplus items back to the sellers who accepted the acquisition offer. The speculator's auction is referred to as the ``return and refund'' auction.

If sellers' items are heterogenous, they will not compete with each other in the return and refund auction, which undermines the speculator's incentive to carry out one. As the purpose of the analysis is to understand whether and by how much the enhanced speculation scheme can improve the speculator's profit, I assume in this part (\Cref{enhanced}) only that the items are homogeneous. Another way to have competition in the return and refund auction is that upon winning the procurement auction, the speculator can conduct the return and refund auction before delivering an item to the auctioneer. In this case, the bidding in the procurement auction must be for a generic item, rather than a specific one.\footnote{For example, if bidders bid for a road pavement contract by promising to finish the job without specifying the identity of the contractor they will use, the bidding is considered to be generic.}

The three-stage SPA-enhanced-speculation game is described as follows. 
The first two stages are the same as described in \Cref{model}. 
In the third stage that happens after the procurement auction, the speculator conducts an auction to sell back his surplus items and get some refunds from the sellers who accepted the acquisition offer. 
%Since we are interested in speculation activities that do not harm efficiency, w
I assume that the speculator's return and refund auction takes the form of a VCG auction. Specifically, 
if the speculator has $K\geq1$ item(s) for sale after the procurement auction, the return and refund auction is a $(K+1)$-st price auction.\footnote{It is clear that this enhanced speculation model has a forward auction counterpart, in which the speculator can promise to sell to more than one potential buyers before a forward auction. After the auction, the speculator conducts another auction to buy back some promises.}$^,$\footnote{The forward auction counterpart of the baseline setting (where return and refund is not feasible) is described as follows. $N\geq2$ buyers, whose values are independently drawn from $[0,1]$, seek to purchase an item in an upcoming auction. In the meantime, an outside market sells an unlimited amount of this item at a price of 1. 
A speculator
%, who has no item for sale and no value for an item, 
may offer to sell to the potential buyers at a price of $p$ before the auction starts. 
More precisely, the speculator offers each potential buyer a future contract. If a buyer accepts, the buyer pays $p$ and the speculator promises to deliver the item \emph{after the auction concludes}. Buyers who agree to buy from the speculator (rationally) abstain from the auction and the speculator participates in the auction if any contract is signed. 
After the auction concludes, the speculator delivers the item(s) won in the auction and/or bought from the outside market as promised.
}

The limited access and seller asymmetry specifications are maintained in this part. Again, I consider a set of prices that can induce a vector of acceptance/rejection cutoffs in equilibrium. For any vector $\bm v^*\equiv(v_j^*)_{j\in\A}$ with $v_j^*\in[0,r]$,\footnote{It can be shown that $v_j^*>r$ is suboptimal for the speculator.}
 define 
\begin{equation}\label{enhanced p}
\begin{split}
\bar p_j^*(\bm v^*):=&\int_0^{v_j^*}\prod_{s\in\A\setminus\{j\}}\left[1-F(\min\{x,v_s^*\})\right]dx+
\\
&~~~~\int_{v_j^*}^r\prod_{i\in\I\setminus\A}[1-F_i(x)]\prod_{s\in\A\setminus\{j\}}\left[1-F(\max\{x,v_s^*\})\right]dx.
\end{split}
\end{equation}
\Cref{enhanced eqm} provides equilibrium characterization for the enhanced speculation game. 

\begin{proposition}\label{enhanced eqm}
In the three-stage SPA-enhanced-speculation game, suppose that the speculator offers $p_j = \bar p_j^*(\bm v^*)$ to seller $j\in\A$, with $v_j^*\in[0,r]$. Then a PBE of the speculation game is described as follows. Seller $j\in\A$ accepts the speculator's offer if and only if $v_j<v_j^*$.  Sellers bid truthfully in the procurement auction or in the return and refund auction. The speculator engages in strategic supply withholding in the procurement auction.
\end{proposition}

From the speculator's perspective, the enhanced speculation scheme has two advantages over the simple speculation scheme. First, 
holding fixed a vector of acceptance/rejection cutoffs $\bm v^*$, 
the compensation to the sellers is lower under the enhanced speculation approach as the sellers have an opportunity to win their item back.\footnote{It is easy to see that
$
p_j^*(\bm v^*)-\bar p_j^*(\bm v^*)=\int_0^{v_j^*}\left\{1-\prod_{s\in\A\setminus\{j\}}\left[1-F(\min\{x,v_s^*\})\right]\right\}dx>0.
$}
Second, the speculator can get some revenue from the return and refund auction. 
\Cref{more profit} ensues as a result.

\begin{corollary}\label{more profit}
The enhanced speculation approach can 
generate more profit for the speculator (than the simple speculation approach).
\end{corollary}

\Cref{spa knock}
demonstrates the full potential of the enhanced speculation scheme in second-price auctions. 
If the speculator is capable of reaching every seller and organizing the return and refund auction, he is in a position to ``take over'' the procurement auction from the auctioneer.  
\begin{corollary}\label{spa knock}
If $\mathcal A = \mathcal I$, the speculator 
can ``knock out'' every seller and achieve the following outcome by setting $v_i^*=r$ for all $i\in\I$: It is as if the speculator conducts a second-price auction with a reserve price of $r$ to buy an item from the sellers, and then sells the item to the auctioneer at a price of $r$.
\end{corollary}

As a comparison, fix $\A=\I$ and
consider enhanced speculation in first-price auctions. 
\Cref{no knock} shows that complete knockout is generally not feasible. This suggests that enhanced speculation is also more profitable in second-price auctions.
\begin{proposition}\label{no knock}
If $F(r)<1$, complete knockout does not work in FPAs even if the speculator can conduct a VCG auction to return and refund the extra items.\footnote{If $F(r)=1$, the speculator can threat to bid 0 if any seller rejects his offer. Then a complete knockout can be supported as a weak PBE. If $F(r)<1$, because rejection happens with a positive probability on the equilibrium path, the threat would violate sequential rationality.} 
\end{proposition}

\section{Conclusion}\label{section conclusion}
This paper studies speculation in procurement auctions by the means of market power accumulation in the pre-auction stage and supply withholding in the auction stage. The PBE of a dynamic speculation game is characterized in both the case of second-price auctions and the case of first-price auctions. 
Speculation tends to be successful in second-price auctions but not so much in first-price auctions, because sellers respond differently in bidding to the speculator under the two formats. 

In particular, first-price auctions can prevent speculation in some cases. Since speculation harms efficiency and inflates the procurement cost, in those cases, first-price auctions would outperform second-price auctions. 

%%%Since speculation harms the auctioneer, first-price auctions would be a better choice in the case where they render speculation unprofitable. In that case, 
%%%
%%%
%%%Speculation has . Because the speculator does not value the items acquired as much as the sellers do, private values are  often lost and efficiency is impaired. 
%%%Sellers benefit from the presence of the speculator, since they are offered a new option of selling to the speculator, in which case they are well-compensated. 
%%%The auctioneer suffers from speculation as the procurement price is driven up. 
%%%
%%%One implication is that first-price auctions would outperform second-price auctions 

\clearpage
\begin{appendices}
\section{Proofs}
\label{proofs}

\begin{description}
\item[Proof of \Cref{spa_cutoff}]
\end{description}
\begin{proof}
I prove the lemma in two steps. In step 1, I show that for each seller $i\in\I$, there exists $v_i^*\in[0,r]$ such that seller $i$ takes the speculator's offer if and only if $v_i<v_i^*$. I then show in step 2 that $v_i^*$s are the same across $i\in\I$. 

\paragraph{Step 1.} Given that all the other sellers and the speculator play their equilibrium strategies, consider seller $i$'s problem.
It is useful to note that if seller $i$ accepts the speculator's offer, his payoff is 
\[
\pi_i^A(v_i) = p-v_i.
\]
Suppose first that all the other sellers reject the speculator's offer with probability 1. Then if seller $i$ rejects the offer, he competes in a second-price auction against all the other sellers. It is straightforward to see that his payoff is 
\[
\pi_i^R(v_i)= \int_{v_i}^r[1-F(x)]^{N-1}\1_{\{x\leq r\}}dx.\footnote{The payoff calculation here follows from a standard payoff equivalence argument. Since in the direct mechanism seller $i$'s winning probability is $[1-F(v_i)]^{N-1}\1_{\{v_i\leq r\}}$, and his payoff is 0 if his value is realized to be $r$, seller $i$'s interim expected payoff can be obtained by integrating the winning probability from $v_i$ to $r$. 
} 
\]
Therefore,
\[
\frac{\partial[\pi_i^A(v_i)-\pi_i^R(v_i)]}{\partial v_i} =\begin{cases}
 -\{1-[1-F(v_i)]^{N-1}\}<0,&\text{ if } 0<v_i\leq r,\\
-1<0,&\text{ if } v_i> r.
  \end{cases}
\]
%That $\frac{\partial[\pi_i^A(v_i)-\pi_i^R(v_i)]}{\partial v_i}<0$
This implies that $\pi_i^A(v_i)-\pi_i^R(v_i)$ strictly decreases with $v_i$, which further implies that there exists a desired cutoff $v_i^*$.

Next, suppose that the probability of at least one other seller taking the speculator's offer, denoted by $\mathcal Q$, is positive. If seller $i$ rejects the offer, his payoff is 0 if at least one other seller accepts the offer, which happens with probability $\mathcal Q$. If no seller accepts the offer, or equivalently, after the public history $(N,0)$, seller $i$'s payoff is 
\[
\pi_i^R(v_i;N,0) = \int_{v_i}^r\mathcal{WP}_i(x;N,0)dx, 
\]
where $\mathcal{WP}_i(x;N,0)$ stands for the winning probability of seller $i$ with a realized value $x$. 
Then seller $i$'s interim expected payoff is 
\[
\pi_i^R(v_i) = (1-\mathcal Q)\int_{v_i}^r\mathcal{WP}_i(x;N,0)dx,
\]
which implies that 
\[
\frac{\partial[\pi_i^A(v_i)-\pi_i^R(v_i)]}{\partial v_i} =-[1-(1-\mathcal Q)\mathcal{WP}_i(v_i;N,0)]<0.
\]
Again, because $\pi_i^A(v_i)-\pi_i^R(v_i)$ strictly decreases with $v_i$, there exists a desired cutoff $v_i^*$. In particular, the following holds true for the cutoff $v_i^*$,
\begin{equation}\label{spa_indiff}
\pi_i^A(v_i^*) \leq \pi_i^R(v_i^*),\text{ and }\pi_i^A(v_i^*) = \pi_i^R(v_i^*)\text{ if }v_i^*>0.
\end{equation}

\paragraph{Step 2.} Without loss of generality, suppose that $v_1^*\leq v_2^*\leq\ldots\leq v_N^*$. Assume by contradiction that $v_\ell^*<v_{\ell+1}^*$ for some $1\leq\ell< N$. Note that for each $i\in\I$, given $v_i\geq v_i^*$, seller $i$'s winning probability is 
\[
\mathcal{WP}_i(v_i)=[1-F(v_i)]^{s-1}\prod_{t=s+1}^N[1-F(v_t^*)]\1_{\{v_i\leq r\}},\text{ if }v_s^*\leq v_i< v_{s+1}^*.
\]
%%Because truthful bidding is optimal for seller $\ell$, his payoff can be calculated as 
Consider seller $\ell$ with value $v_\ell^*$. If he rejects the acquisition offer, his expected payoff from the auction is 
\[\pi_\ell^R(v_\ell^*)=\int_{v_\ell^*}^r\mathcal{WP}_\ell(x)dx.\] 
It follows from \eqref{spa_indiff} that
\begin{equation}\label{cutoff ineq}
p-v_\ell^* \leq \pi_\ell^R(v_\ell^*) =  \int_{v_\ell^*}^{v_{\ell+1}^*}[1-F(x)]^{\ell-1}\prod_{t=\ell+1}^N[1-F(v_t^*)]\1_{\{v_\ell\leq r\}}dx+\int_{v_{\ell+1}^*}^r\mathcal{WP}_\ell(x)dx.
\end{equation}
Since 
\[
[1-F(x)]^{\ell-1}\prod_{t=\ell+1}^N[1-F(v_t^*)]\1_{\{v_\ell\leq r\}}\leq 1-F(v_{\ell+1}^*)<1, 
\]
\eqref{cutoff ineq} implies that
\begin{equation}\label{eq1}
p-v_\ell^* <v_{\ell+1}^*-v_{\ell}^*+\int_{v_{\ell+1}^*}^r\mathcal{WP}_\ell(x)dx.
\end{equation}

Because $v_{\ell+1}^*>0$, one can see that 
\begin{equation}\label{eq2}
p-v_{\ell+1}^* =\int_{v_{\ell+1}^*}^r\mathcal{WP}_{\ell+1}(x)dx = \int_{v_{\ell+1}^*}^r\mathcal{WP}_{\ell}(x)dx,
\end{equation}
where the the first equality follows from \eqref{spa_indiff} and the second equality is due to the fact that $\mathcal{WP}_{\ell}(x) = \mathcal{WP}_{\ell+1}(x)$ for $x\geq v_{\ell+1}^*$. It is easy to see that a contradiction arises between \eqref{eq1} and \eqref{eq2}. This completes the proof. 
\end{proof}

\begin{description}
\item[Proof of \Cref{spa_eqm}]
\end{description}
\begin{proof}
By \Cref{spa_cutoff}, it suffices to show that the equilibrium cutoff is unique and is given by $v^*(p)$. 

If seller $i$ rejects the speculator's offer, it is weakly dominant for him to bid truthfully in the procurement auction. Given that other sellers follow the equilibrium strategy, the winning probability of seller $i$ is 
\[
\mathcal{WP}_i(v_i) = \begin{cases}
[1-F(v^*)]^{N-1}\1_{\{v_i\leq r\}},&\text{ if }v_i<v^*,\\
[1-F(v_i)]^{N-1}\1_{\{v_i\leq r\}},&\text{ if }v_i\geq v^*.\\
\end{cases}
\]
Then \eqref{spa_indiff} in the proof of \Cref{spa_eqm} can be rewritten as follows, 
\[
p-v^*\leq \int_{v^*}^r [1-F(x)]^{N-1}\1_{\{x\leq r\}}dx,\text{ and }p-v^*= \int_{v^*}^r [1-F(x)]^{N-1}\1_{\{x\leq r\}}dx\text{ if }v^*>0.
\]
This immediately implies that $v^*=v^*(p)$.
\end{proof}

\begin{description}
\item[Proof of \Cref{spa_profitable}]
\end{description}
\begin{proof}
Note that $\Pi^*(v^*)$ can be rewritten as follows. 
\begin{align*}
\Pi^*(v^*)
%:&= \sum_{m=0}^{N-1}{N\choose m}[F(v^*)]^{N-m}[1-F(v^*)]^my(m,v^*)-NF(v^*)p^*(v^*)\\
&=\sum_{m=0}^{N-2}{N\choose m}[F(v^*)]^{N-m}[1-F(v^*)]^m\left\{v^*+\int_{v^*}^r\left[\frac{1-F(x)}{1-F(v^*)}\right]^mdx\right\}\\
&~~~~- NF(v^*)\left\{1-[1-F(v^*)]^{N-1}\right\}v^*.
\end{align*}
As $v^*\to0$, the only non-zero term of $\lim_{v^*\to0}\frac{\Pi^*(v^*)}{[F(v^*)]^2}$ is 
\[
\lim_{v^*\to0}{N\choose {N-2}}[1-F(v^*)]^{N-2}\int_{v^*}^r\left[\frac{1-F(x)}{1-F(v^*)}\right]^{N-2}dx = {N\choose 2}\int_0^r[1-F(x)]^{N-2}dx>0.
\]

Since $\frac{\Pi^*(v^*)}{[F(v^*)]^2}$ is continuous in $v^*\in[0,r]$, $\lim_{v^*\to0}\frac{\Pi^*(v^*)}{[F(v^*)]^2}>0$ implies that $\Pi^*(v^*)>0$ for small enough $v^*>0$. Consequently, the speculator's expected profit is 
$\max_{v^*\in[0,r]}\Pi^*(v^*)>0$. This concludes the proof. 
\end{proof}

\begin{description}
\item[Proof of \Cref{fpa cutoff}]
\end{description}
\begin{proof} Let $\alpha(v)$ denote the probability that a type-$v$ seller accepts the speculator's offer in equilibrium. 
Define $\ubar v: = \inf\{v\in[0,1]: \exists \epsilon>0\text{ such that }\alpha(v')<1\text{ for all }v'\in\mathcal U_\epsilon(v):=(v,v+\epsilon)\}$. By construction, $\alpha(v) =1$ for all $v<\ubar v$ except for a set with $F$-measure 0. It is clear that $\ubar v\leq p\leq r$. 

It suffices to show that $\alpha(v) = 0$ for all $v>\ubar v.$
Consider seller $i$ with $v_i>\ubar v$. 
If seller $i$ accepts the speculator's offer, his payoff is  
\[
\varpi^A(v_i) = p-v_i.
\]
If seller $i$ rejects the speculator's offer, the interim expected payoff to seller $i$ can be established as follows by a standard payoff equivalence argument,  
\[
\varpi^R(v_i) = \varpi^R(\ubar v)-
\sum_{m=1}^{N-1}\pr[h = (m,1)]\int_{\footnotesize\ubar v}^{v_i}\mathcal{WP}^\star(x;m,1)dx-\pr[h = (N,0)]\int_{\footnotesize\ubar v}^{v_i}\mathcal{WP}^\star(x;N,0)dx,
\]
where $\mathcal{WP}^\star(x;h)$ is the winning probability of seller $i$ with value $x$ after a history $h$, given that all the other bidders in the auction use their equilibrium bidding strategies and seller $i$ best-responds to the other bidders' equilibrium bidding. Note that 
\[
\frac{\partial\varpi^R(v_i)}{\partial v_i} = -\sum_{m=1}^{N-1}\pr[h = (m,1)]\mathcal{WP}^\star(v_i;m,1)-\pr[h = (N,0)]\mathcal{WP}^\star(v_i;N,0)\geq -1.
\]
By the definition of $\ubar v$, one can see that $\alpha(v_i)<1$ for $v_i\in \mathcal U_\epsilon(\ubar v)$. Therefore, 
 $\pr[h = (N,0)]>0$ and $\mathcal{WP}^\star(v_i;N,0)<1$ for $v_i\in \mathcal U_\epsilon(\ubar v)$. This implies that $\frac{\partial\varpi^R(v_i)}{\partial v_i}>-1$ for $v_i\in \mathcal U_\epsilon(\ubar v)$. Further,
\begin{equation}\label{rej ineq}
\varpi^R(v_i) = \varpi^R(\ubar v)+\int_{\footnotesize\ubar v}^{v_i}\frac{\partial\varpi^R(x)}{\partial x}dx> \varpi^R(\ubar v)-(v_i-\ubar v).
\end{equation}
Since the speculator's offer is rejected with positive probability when a seller's value is in $\mathcal U_\epsilon(\ubar v)$, we know that 
\[
\varpi^R(\ubar v)\geq \varpi^A(\ubar v) = p-\ubar v.
\]
Plugging this into \eqref{rej ineq} yields that 
$
\varpi^R(v_i)>p-v_i = \varpi^A(v_i),
$
which implies that $\alpha(v_i) = 0$. This completes the proof. 
\end{proof}

\begin{description}
\item[Proof of \Cref{fpa payoff}]
\end{description}
\begin{proof}
Suppose that the sellers' equilibrium bidding strategy is given by $\beta(v;m,v^\star)$ for $v\in[v^\star,1]$. 
Define the distribution of a seller's equilibrium bid as $H(b):=\pr[\beta(v;m,v^\star)\leq b]$.
Denote the speculator's lowest and highest equilibrium bid by $\ubar b$ and $\bar b$, respectively.\footnote{More precisely, $\bar b:=\inf\{b'\geq0:\pr[\text{the speculator's bid in equilibrium}\leq b']=1\}$, and $\ubar b:=\sup\{b'\geq0:\pr[\text{the speculator's bid in equilibrium}\geq b']=1\}$.} Clearly, $v^\star\leq\ubar b\leq \bar b\leq r$.
I first establish the following facts. 

\begin{claim}\label{fact1}
$H(b)$ has no mass point at $b$ for an arbitrary $b\leq\bar b$. 
\end{claim}
\begin{proof}
Suppose to the contrary that $\pr[\beta(v;m,v^\star)=b']>0$ for some $b'\leq \bar b$. Then there exists $v'<b'$ such that $\beta(v';m,v^\star) = b'$. A type-$v'$ seller would have an incentive to deviate to bidding $b'-\epsilon$ for some small $\epsilon>0$ to significantly improve his winning probability.
\end{proof}

\begin{claim}\label{fact2}
$H(\bar b) = G(\bar b;v^\star)$.
\end{claim}
\begin{proof}
 It suffices to show that $\beta(v;m,v^\star)<\bar b$ for all $v<\bar b$. Suppose to the contrary that $\beta(v';m,v^\star)\geq\bar b$ for some $v'<\bar b$. Then a type-$v'$ seller loses for sure. He can profitably deviate to bidding $\bar b-\epsilon$ for some small $\epsilon>0$. 
 \end{proof}

\begin{claim}\label{fact3}
$\beta(v;m,v^\star)$ is strictly increasing in $v\in(v^\star,\bar b)$. 
\end{claim}
\begin{proof}
Suppose $v^\star<v'<v''<\bar b$. The no-deviation condition of the two types are 
\begin{align*}
[\beta(v';m,v^\star)-v']\mathcal{WP}(\beta(v';m,v^\star))&\geq[\beta(v'';m,v^\star)-v']\mathcal{WP}(\beta(v'';m,v^\star)),\\
[\beta(v'';m,v^\star)-v'']\mathcal{WP}(\beta(v'';m,v^\star))&\geq[\beta(v';m,v^\star)-v'']\mathcal{WP}(\beta(v';m,v^\star)),
\end{align*}
where $\mathcal{WP}(b)$ stands for the winning probability of bidding $b$ given that other players use their equilibrium bidding strategies. Adding up the two no-deviation conditions yields the following,
\[
(v''-v')[\mathcal{WP}(\beta(v';m,v^\star))-\mathcal{WP}(\beta(v'';m,v^\star))]\geq0,
\]
which implies that $\mathcal{WP}(\beta(v';m,v^\star))\geq\mathcal{WP}(\beta(v'';m,v^\star))$ and thus $\beta(v';m,v^\star)\leq \beta(v'';m,v^\star)$. Since $H(b)$ has no mass point at $b$ for an arbitrary $b\leq\bar b$ and $\beta(v;m,v^\star)<\bar b$ for $v<\bar b$, $\beta(v;m,v^\star)$ is strictly increasing in $v\in(v^\star,\bar b)$.
\end{proof}

\begin{claim}\label{fact4}
$\beta(v^\star;m,v^\star) = \ubar b$ and thus $H(\ubar b) = 0$. 
\end{claim}
\begin{proof}
If $\beta(v^\star;m,v^\star) > \ubar b$, the speculator can profitably deviate from bidding $\ubar b$ to bidding $\beta(v^\star;m,v^\star)$, so $\beta(v^\star;m,v^\star) \leq \ubar b$. Suppose by contradiction that $\beta(v^\star;m,v^\star) < \ubar b$. 
%%Note that $\beta(v;m,v^\star)$ increases continuously when $\beta(v;m,v^\star)\leq\ubar b$. Otherwise if a jump exists---i.e., $\beta(v+0;m,v^\star)>\beta(v-0;m,v^\star)$ for some $v$ satisfying $\beta(v;m,v^\star)\leq\ubar b$---the $v-0$ type can profitably deviate to bidding $\beta(v-0;m,v^\star)+\epsilon$ for a small enough $\epsilon>0$.
Then there exists $\hat v>v^\star$ such that $\beta(\hat v;m,v^\star)=\ubar b$. Note that for $v\leq \hat v$, the speculator's payoff from bidding $\beta(v;m,v^\star)$ is $[1-G(v;v^\star)]^m\beta(v;m,v^\star)$. I show next that $[1-G(v;v^\star)]^m\beta(v;m,v^\star)$ is decreasing in $v\in[v^\star,\hat v]$, which implies that the speculator can profitably deviate from bidding $\ubar b$ to bidding $\beta(v^\star;m,v^\star)$.

By the payoff equivalence theorem, for $v\in[v^\star,\hat v]$, 
\[
(\ubar b-\hat v)[1-G(\hat v;v^\star)]^{m-1}+\int_v^{\hat v}[1-G(x;v^\star)]^{m-1}dx = [1-G(v;v^\star)]^{m-1}[\beta(v;m,v^\star)-v].
\]
Therefore, 
\[
\frac{d\left\{ [1-G(v;v^\star)]^{m-1}\beta(v;m,v^\star)\right\}}{dv} = v\frac{d[1-G(v;v^\star)]^{m-1}}{dv} <0.
\]
This implies that $[1-G(v;v^\star)]^{m-1}\beta(v;m,v^\star)$ decreases in $v\in[v^\star,\hat v]$. Clearly, $[1-G(v;v^\star)]^{m}\beta(v;m,v^\star)$ decreases in $v\in[v^\star,\hat v]$.
\end{proof}

I then proceed to prove the lemma. Note that bidding $\ubar b$ is optimal for the speculator. By \Cref{fact4}, the speculator's payoff from bidding $\ubar b$ is $\ubar b$, as he wins the auction with probability 1. 
Therefore, for all $b\in[v^\star,r]$, it holds that
\begin{equation}\label{ineq1}
\begin{split}
\ubar b&\geq b\times\pr[\text{The speculator wins with a bid of $b$}]\\
&\geq b\times\pr[\text{The speculator wins with a bid of $b$ and without a tie}]\\
&= b\left\{1-\Pr[\beta(v;m,v^\star)\leq b]\right\}^m\\
&\geq b[1-G(b;v^\star)]^m,
\end{split}
\end{equation}
where the first inequality follows from the optimality of bidding $\ubar b$, and the last inequality follows from $\beta(v;m,v^\star)\leq v$. By \Cref{fact1} and \Cref{fact2}, the speculator's payoff from bidding $\bar b$ is $\bar b[1-G(\bar b;v^\star)]^m$. Due to the optimality of bidding $\bar b$ and \eqref{ineq1}, it holds that 
\[
\bar b[1-G(\bar b;v^\star)]^m = \ubar b\geq b[1-G(b;v^\star)]^m\text{ for all }b\in[v^\star,r].
\]
This implies that $\ubar b = \max_{r\geq b\geq v^\star}b[1-G(b;v^\star)]^m \equiv \ubar b(m,v^\star)$. Consequently, the speculator's equilibrium payoff is $\ubar b(m,v^\star)$.

Finally, consider the interim equilibrium payoff of a type-$v^\star$ seller. If the speculator bids $\ubar b$ with probability 0, by \Cref{fact4} the seller's interim equilibrium payoff is $\ubar b(m,v^\star)-v^\star$. If the speculator bids $\ubar b$ with a positive probability, then $\ubar b = v^\star$. Otherwise the type-$v^\star$ seller would have an incentive to deviate to bidding $\ubar b-\epsilon$ for some small $\epsilon>0$. Plugging $\ubar b = v^\star$ into \eqref{ineq1} yields that 
\[
v^\star \geq b[1-G(b;v^\star)]^m\text{ for all }b\in[v^\star,r].
\]
As a result, $\ubar b(m,v^\star) = v^\star$. Then it is clear that the interim equilibrium payoff of the seller is $0=\ubar b(m,v^\star)-v^\star$.
This completes the proof.
\end{proof}

\begin{description}
\item[Proof of \Cref{fpa sub}]
\end{description}
\begin{proof}
First, consider the speculator's problem, holding fixed the sellers' equilibrium bidding. The speculator's expected payoff from bidding $b\in[0,r]$ is
\[
\begin{cases}
b,&\text{ if }b\leq \ubar b(m,v^\star),\\
b\{\pr[\beta(v;m,v^\star)>b]\}^m,&\text{ if }\ubar b(m,v^\star)<b\leq r,\\
0,&\text{ if }r< b.
\end{cases}
\]
Clearly, the speculator would not deviate to bidding $b<\ubar b(m,v^\star)$. 
Note that 
%%%\[
%%%\begin{split}
%%%b\{\pr[\beta(v;m,v^\star)>b]\}^m &= b\left\{\pr\left[G(v;v^\star)>1-\left[\frac{\ubar b(m,v^\star)}{b}\right]^{1/m}\right]\right\}^m
%%%\\&=
%%%\begin{cases}
%%%\ubar b(m,v^\star),&\text{ if }b\leq \bar b(m,v^\star),\\
%%%b[1-G(b;v^\star)]^m,&\text{ if }b> \bar b(m,v^\star).
%%%\end{cases}
%%%\end{split}
%%%\]
\[
b\{\pr[\beta(v;m,v^\star)>b]\}^m =\begin{cases}
\ubar b(m,v^\star),&\text{ if }b\leq \bar b(m,v^\star),\\
b[1-G(b;v^\star)]^m,&\text{ if }\bar b(m,v^\star)<b\leq r.
\end{cases}
\]
By the definition of $\ubar b(m,v^\star)$, $\ubar b(m,v^\star)\geq b[1-G(b;v^\star)]^m$ for all $b\in[v^\star,r]$, so the speculator would not deviate to bidding $b>\bar b(m,v^\star)$. Bidding anything in between $\ubar b(m,v^\star)$ and $\bar b(m,v^\star)$ is optimal for the speculator. 

Next, consider the problem of seller $i\in\I$, holding fixed all the other sellers' equilibrium bidding and the speculator's. If $v_i\geq\bar b(m,v^\star)$, seller $j$ can never win the auction with a positive payoff, bidding truthfully yields a zero payoff and thus is optimal for him.

Suppose $v_i\in[v^\star,\bar b(m,v^\star))$. Seller $j$ never bids strictly less than $\ubar b(m,v^\star)$. Otherwise, he can deviate to a slightly higher bid and still win for sure. Bidding above $\bar b(m,v^\star)$ is also not optimal for seller $j$ since he loses for sure in that case.  Therefore, seller $j$ chooses a bid in $[\ubar b(m,v^\star),\bar b(m,v^\star)]$ to maximize his expected payoff. This is equivalent to choosing $\hat v \in[v^\star,\bar b(m,v^\star)]$ and bidding $\beta(\hat v;m,v^\star)$. Formally, seller $j$ solves
\[
\max_{\hat v\in[v^\star,\bar b(m,v^\star)]}\varpi(\hat v;v) := [1-G(\hat v;v^\star)]^{m-1}[1-\Psi(\beta(\hat v;m,v^\star))][\beta(\hat v;m,v^\star)-v].
\]
It can be verified that 
\[
\frac{\partial \varpi(\hat v;v)}{\partial \hat v} =\frac{m\beta(\hat v;m,v^\star)}{\beta(\hat v;m,v^\star)-\hat v}[1-G(\hat v;v^\star)]^{m-2}G'(\hat v;v^\star)[1-\Psi(\beta(\hat v;m,v^\star))](v-\hat v).
\]
It follows from the definition of $\ubar b(m,v^\star)$ that $\beta(v;m,v^\star)> v$ for all $v\in[v^\star,\bar b(m,v^\star))$.
Therefore, $\varpi(\hat v;v)$ is maximized by $\hat v = v$ and it is optimal for seller $j$ to bid $\beta(v;m,v^\star)$.
\end{proof}

\begin{description}
\item[Proof of \Cref{fpa pv}]
\end{description}
\begin{proof}
Consider the right derivative of $p^\star(v)$ for any $v\in(0,r)$. For notational ease, define
\begin{equation}\label{ub der ineq}
\mathcal C(v;m):=[1-F(v)]^m[F(v)]^{N-1-m}[\ubar b(m+1,v)-v].
\end{equation}
I first prove that $\mathcal C_+'(v;m):=\lim_{\epsilon\to0^+}\frac{\mathcal C(v+\epsilon;m)-\mathcal C(v;m)}{\epsilon}>-[1-F(v)]^m[F(v)]^{N-1-m}$.

By definition, \[\ubar b(m+1,v)=\frac{\max_{r\geq b\geq v}b[1-F(b)]^{m+1}}{[1-F(v)]^{m+1}}.\]
If there exists $b'\in(v,r]$ such that $b'\in\text{arg}\max_{r\geq b\geq v}b[1-F(b)]^{m+1}$, then increasing $v$ marginally would not change the value of $\max_{r\geq b\geq v}b[1-F(b)]^{m+1}$. It follows that
\[
\begin{split}
\mathcal C_+'(v;m)&= \frac{\partial \left\{
[F(v)]^{N-1-m}b'[1-F(b')]^{m+1}/[1-F(v)]
\right\}}{\partial v}
-\frac{\partial \left\{[1-F(v)]^m[F(v)]^{N-1-m}v\right\}}{\partial v}\\
&=b'[1-F(b')]^{m+1}\frac{
(N-1-m)[F(v)]^{N-2-m}[1-F(v)]+[F(v)]^{N-1-m}
}{[1-F(v)]^2}f(v)\\
&
~~~~  -v\left\{
(N-1-m)[F(v)]^{N-2-m}[1-F(v)]^{m}-m[1-F(v)]^{m-1}[F(v)]^{N-1-m}
\right\}f(v)\\
&
~~~~  -[1-F(v)]^m[F(v)]^{N-1-m}
\\
&>-[1-F(v)]^m[F(v)]^{N-1-m},
\end{split}
\]
where the inequality follows from the fact that $b'[1-F(b')]^{m+1}\geq v[1-F(v)]^{m+1}$.

If $v[1-F(v)]^{m+1}>b[1-F(b)]^{m+1}$ for all $b\in(v,r]$, there exists $\epsilon>0$ such that $\ubar b(m+1,v') = v'$ for all $v'\in[v,v+\epsilon]$. As a result, $\mathcal C(v';m) = 0$ for $v'\in[v,v+\epsilon]$. This implies that
\[
\mathcal C_+'(v;m)= 0>-[1-F(v)]^m[F(v)]^{N-1-m}.
\]

To sum up, \eqref{ub der ineq} holds and the following ensues,
\[
\begin{split}
\lim_{\epsilon\to0^+}\frac{p^\star(v+\epsilon)-p^\star(v)}{\epsilon} &= 1-[1-F(v)]^{N-1}+\sum_{m=0}^{N-2}\binom{N-1}{m}\mathcal C_+'(v;m)\\
&>1-[1-F(v)]^{N-1}-\sum_{m=0}^{N-2}\binom{N-1}{m}[1-F(v)]^m[F(v)]^{N-1-m} = 0.
\end{split}
\]
Consequently, $p^\star(v)$ is strictly increasing in $v\in(0,r)$. 

Finally, it is easy to verify that $p^\star(0) = \pi_0$ and $p^\star(r) = r$. 
This completes the proof. 
\end{proof}

\begin{description}
\item[Proof of \Cref{fpa eqm}]
\end{description}
\begin{proof}
Point (i) follows immediately from \Cref{fpa cutoff} and the indifference condition \eqref{fpa indiff}.

To prove Point (ii), it suffices to verify the cutoff acceptance strategy is optimal for a representative seller $i\in\I$. If seller $i$, with a realized value of $v_i$, accepts the acquisition offer, his payoff is
\[
\varpi_i^A(v_i) = p-v_i.
\]
If seller $i$ rejects the acquisition offer, he participates in the procurement auction. As is shown in the proof of \Cref{fpa cutoff}, seller $i$'s expected payoff $\varpi_i^R(v_i)$ satisfies 
\[
\frac{\partial \varpi_i^R(v_i)}{\partial v_i}\geq-1.
\]
As a result, $\varpi_i^A(v_i)-\varpi_i^R(v_i)$ decreases with $v_i$. Since $\varpi_i^A(v^\star)=\varpi_i^R(v^\star)$, it is optimal for seller $i$ to accept the acquisition offer if and only if $v_i<v^\star$.
\end{proof}

\begin{description}
\item[Proof of \Cref{fpa unprof}]
\end{description}
\begin{proof}
See \Cref{fpa num} in the main text. 
\end{proof}

\begin{lemma}\label{auction_payment}
For all $v^\dagger\in[0,r)$,
\[
y(m,v^\dagger):=v^\dagger+\int_{v^\dagger}^r\left[\frac{1-F(x)}{1-F(v^\dagger)}\right]^mdx>\ubar b(m,v^\dagger):=\max_{r\geq b\geq v^\dagger}b[1-G(b;v^\dagger)]^m.
\]
\end{lemma}

\begin{description}
\item[Proof of \Cref{auction_payment}]
\end{description}
\begin{proof}
Recall that $G(v;v^\dagger):=\frac{F(v)-F(v^\dagger)}{1-F(v^\dagger)}$ for $v\in[v^\dagger, 1]$. So it suffices to prove that
\[
v^\dagger+\int_{v^\dagger}^r\left[1-G(x;v^\dagger)\right]^mdx>b[1-G(b;v^\dagger)]^m\text{ for all }v^\dagger\in[0,r)\text{ and }b\in[v^\dagger,r].
\]
In fact, 
\begin{align*}
v^\dagger+\int_{v^\dagger}^r\left[1-G(x;v^\dagger)\right]^mdx&\geq
v^\dagger+\int_{v^\dagger}^b\left[1-G(x;v^\dagger)\right]^mdx\\
&\geq v^\dagger+\int_{v^\dagger}^b\left[1-G(b;v^\dagger)\right]^mdx\\
&=v^\dagger\left\{1-\left[1-G(b;v^\dagger)\right]^m\right\} + b[1-G(b;v^\dagger)]^m\\
&\geq b[1-G(b;v^\dagger)]^m.
\end{align*}
The equalities cannot hold at the same time, as that requires
 $b=r$ (from the first inequality), $G(x;v^\dagger) = G(b;v^\dagger)$ for almost every $x\in[v^\dagger,b]$ (from the second inequality), and $v^\dagger =0$ (from the last inequality). 
\end{proof}

\begin{description}
\item[Proof of \Cref{aggressive}]
\end{description}
\begin{proof}
It suffices to prove that $\tilde\beta(v;m,v^\star)>\beta(v;m,v^\star)$ for all $v\in[v^\star,\bar b(m,v^\star))$, which is equivalent to
\[
\mathcal{Z}(v):=v[1-G(v;v^\star)]^m+\int_v^r[1-G(x;v^\star)]^mdx>\ubar b(m,v^\star)\text{ for all }v\in[v^\star,\bar b(m,v^\star)).
\]
It is easy to verify that $\mathcal{Z}(v)$ is strictly decreasing in $v\in[v^\star,\bar b(m,v^\star)]$. Then the proof is completed by the following fact,
\[\mathcal{Z}(\bar b(m,v^\star))=\ubar b(m,v^\star) +\int_{\bar b(m,v^\star)}^r[1-G(x;v^\star)]^mdx\geq \ubar b(m,v^\star).\]
\end{proof}

\begin{description}
\item[Proof of \Cref{spa eqm 2}]
\end{description}
\begin{proof}
It suffices to verify that the acceptance/rejection schedule is optimal for an arbitrary seller $j\in\A$, given that all the other sellers follow the equilibrium strategies described in \Cref{spa eqm 2}. 

If seller $j$, with a realized valuation $v_j\leq r$,\footnote{Note that $p_j^*(\bm v^*)\leq r$, so seller $j$ would reject the acquisition offer if $v_j>r$.} accepts the offer, his payoff is $\pi_j^A(v_j) = p_j(\bm v^*)-v_j$. If he rejects the offer, he participates in the procurement auction. 
Seller $j$ wins in the procurement auction if and only if all the other sellers' valuations are above $v_j$ and the speculator failed to acquire any item.
% which means that for all $s\in\A\setminus\{j\}$, $v_s\geq v_s^*$. 
Formally, seller $j$'s winning probability is 
\[
\mathcal{WP}_j(v_j) = \prod_{i\in\I\setminus\A}[1-F_i(v_j)]\prod_{s\in\A\setminus\{j\}}\left[1-F(\max\{v_j,v_s^*\})\right].
\]
As a result, seller $j$'s expected payoff in the auction is 
$
\pi_j^R(v_j) = \int_{v_j}^r\mathcal{WP}_j(x)dx.
$

The following two facts in combination verify the optimality of accepting the acquisition offer if and only if $v_j<v_j^*$:
\begin{align*}
\frac{\partial[\pi_j^A(v_j)-\pi_j^R(v_j)]}{\partial v_j} = -[1-\mathcal{WP}_j(v_j)]\leq0,\text{ and }\pi_j^A(v_j^*)=\pi_j^R(v_j^*).
\end{align*}
\end{proof}

\begin{description}
\item[Proof of \Cref{asym profit}]
\end{description}
\begin{proof}
I show in what follows that the speculator can obtain a positive expected profit by setting $v_j^*=v^*$ for all $j\in\A$ and appropriately choosing $v^*$. 

Given that $v_j^*=v^*$ for all $j\in\A$, the speculator's profit from the events in which only one seller accepts the offer is
\[
\begin{split}
\Pi^*(v^*,1) &= \sum_{j\in\A}F_{j}(v^*)\prod_{j'\in\A\setminus\{j\}}[1-F_{j'}(v^*)]
\Bigg\{
\int_0^{v^*}\prod_{i\in\I\setminus\A}[1-F_i(x)]dx
\\
&~~~~+\int_{v^*}^r
\frac{\prod_{i\in\I\setminus\{j\}}[1-F_i(x)]}{\prod_{j'\in\A\setminus\{j\}}[1-F_{j'}(v^*)]}
dx
-v^*-\int_{v^*}^r\prod_{i\in\I\setminus\{j\}}[1-F_i(x)]dx
\Bigg\}.
\end{split}
\]
As $v^*$ approaches 0, one can see that\footnote{I use $O(\cdot)$ to denote the same order infinitesimal  and $o(\cdot)$ to denote higher order infinitesimal as $v^*\to0$. That is, for an arbitrary function of $v^*$ denoted by $T(v^*)$, 
$\lim_{v^*\to0} O(T(v^*))/T(v^*)$ equals a non-zero constant, and $\lim_{v^*\to0} o(T(v^*))/T(v^*) = 0$.
}
\[
\begin{split}
\Pi^*(v^*,1) &= \sum_{j\in\A}\sum_{j'\in\A\setminus\{j\}}F_j(v^*)F_{j'}(v^*)\int_{v^*}^r\prod_{i\in\I\setminus\{j\}}[1-F_i(x)]dx\\
&~~~~+\sum_{j\in\A,i\in\I\setminus\A}O\left(v^*F_{j}(v^*)F_{i}(v^*)\right)+\sum_{j\in\A,j'\in\A\setminus\{j\}}o\left(F_{j}(v^*)F_{j'}(v^*)\right)
\\
&=\sum_{j\in\A,j'\in\A\setminus\{j\}}F_j(v^*)F_{j'}(v^*)\left\{\int_{v^*}^r\prod_{i\in\I\setminus\{j\}}[1-F_i(x)]dx+\int_{v^*}^r\prod_{i\in\I\setminus\{j'\}}[1-F_i(x)]dx\right\}\\
&~~~~+\sum_{j\in\A,i\in\I\setminus\A}O\left(v^*F_{j}(v^*)F_{i}(v^*)\right)+\sum_{j\in\A,j'\in\A\setminus\{j\}}o\left(F_{j}(v^*)F_{j'}(v^*)\right).
\end{split}
\]

The speculator's expected profit from the events in which two sellers accept the offers is
\[
\begin{split}
&\Pi^*(v^*,2) = \sum_{j\in\A,j'\in\A\setminus\{j\}}F_{j}(v^*)F_{j'}(v^*)\prod_{j''\in\A\setminus\{j,j'\}}[1-F_{j''}(v^*)]\Bigg\{
\int_0^{v^*}\prod_{i\in\I\setminus\A}[1-F_i(x)]dx
\\
&+\int_{v^*}^r
\frac{\prod_{i\in\I\setminus\{j,j'\}}[1-F_i(x)]}{\prod_{j''\in\A\setminus\{j,j'\}}[1-F_{j''}(v^*)]}
dx
-2v^*-\int_{v^*}^r\prod_{i\in\I\setminus\{j,j'\}}[1-F_i(x)][2-F_j(x)-F_{j'}(x)]dx
\Bigg\}.
\end{split}
\]
As $v^*$ approaches 0, one can see that
\[
\begin{split}
\Pi^*(v^*,2) &= \sum_{j\in\A,j'\in\A\setminus\{j\}}F_{j}(v^*)F_{j'}(v^*)\int_{v^*}^r\prod_{i\in\I\setminus\{j,j'\}}[1-F_i(x)][F_j(x)+F_{j'}(x)-1]dx
\\&~~~~+\sum_{j\in\A,j'\in\A\setminus\{j\}}o\left(F_{j}(v^*)F_{j'}(v^*)\right).
\end{split}
\]

Analogously, one can calculate $\Pi^*(v^*,m)$ for $m\geq3$. For $m\geq3$, it is clear that $\Pi^*(v^*,m) = \sum_{j\in\A}\sum_{j'\in\A\setminus\{j\}}o\left(F_{j}(v^*)F_{j'}(v^*)\right)$ as $v^*$ approaches 0.
Since the speculator's expected profit is 
$
\Pi^*(v^*) = \sum_m \Pi^*(v^*,m), 
$
as $v^*$ approaches 0,\footnote{Note that $|\A|>1$ is needed to make sure $\A\setminus\{j\}$ is not an empty set.} 
\[
\begin{split}
\Pi^*(v^*)  &= \sum_{j\in\A}\sum_{j'\in\A\setminus\{j\}}F_{j}(v^*)F_{j'}(v^*)\int_{v^*}^r\prod_{i\in\I\setminus\{j,j'\}}[1-F_i(x)]dx\\
&~~~~+\sum_{j\in\A,i\in\I\setminus\A}O\left(v^*F_{j}(v^*)F_{i}(v^*)\right)+\sum_{j\in\A,j'\in\A\setminus\{j\}}o\left(F_{j}(v^*)F_{j'}(v^*)\right).
\end{split}
\]

One can find $j_1,j_2\in\A$ such that $\lim_{v\to0}[F_{j}(v)/F_{j_1}(v)] < \infty$ for all $j\in\A$ and that $\lim_{v\to0}[F_{j'}(v)/F_{j_2}(v)] < \infty$ for all $j'\in\A\setminus\{j_1\}$. Then $j_1$ and $j_2$ (as $k$ and $k'$) must satisfy \Cref{con1}. From \Cref{con1} and the specification of $j_1$ and $j_2$, it follows that 
\[
\lim_{v\to0}\frac{vF_{j}(v)F_{i}(v)}{F_{j_1}(v)F_{j_2}(v)} = 0\text{ and }\lim_{v\to0}\frac{
o\left(F_{j}(v^*)F_{j'}(v^*)\right)
}{F_{j_1}(v)F_{j_2}(v)} = 0,\forall j\in\A,j'\in\A\setminus\{j\},\text{ and } i\in\I\setminus\A.
\]
Therefore,
\[
\lim_{v^*\to0}\frac{\Pi^*(v^*)}{F_{j_1}(v^*)F_{j_2}(v^*)}\geq\int_{0}^r\prod_{i\in\I\setminus\{j_1,j_2\}}[1-F_i(x)]dx>0.
\]
This completes the proof. 
\end{proof}

\begin{description}
\item[Proof of \Cref{enhanced eqm}]
\end{description}
\begin{proof}
First, it is useful to note that a seller can either participate in the procurement auction, or the return and refund auction, but not both. Since there are no continuation games for the sellers in the two auctions, it is easy to verify that the bidding strategies described in \Cref{enhanced eqm} are optimal for the sellers. 

For the speculator, the return and refund auction happens after the procurement auction. Therefore, his bidding in the procurement auction may affect how many items he has for sale in the return and refund auction. Suppose that the speculator acquired $K\geq1$ items from $K$ sellers. 
The speculator's opportunity cost for selling one item in the procurement auction is that he can bring one less item to the return and refund auction. Because bringing all $K$ items to the return and refund auction yields a zero auction revenue, the speculator's opportunity cost for selling one item is 0. Therefore, strategic supply withholding---bidding 0 for one item and withholding the rest---is optimal for the speculator. 

It only remains to verify that the acceptance/rejection schedule is optimal for an arbitrary seller $j\in\A$, given that all the other sellers follow their strategies described in \Cref{enhanced eqm}. If seller $j\in\A$, with a realized valuation $v_j\in[0,1]$, accepts the acquisition offer, he gets a payment of $\bar p_j^*(\bm v^*)$ and a chance to participate in the return and refund auction. In the return and refund auction, seller $j$ wins if his valuation is not the lowest among the sellers who accepted the acquisition offer. The winning probability is 
\[
\overline{\mathcal{WP}}_j(v_j) = 1-\prod_{s\in\A\setminus\{j\}}\left[1-F(\min\{v_j,v_s^*\})\right].
\]
Therefore, seller $j$'s expected payoff in the return and refund auction is $\int_0^{v_j}\overline{\mathcal{WP}}_j(x)dx$. Further, seller $j$'s expected payoff from accepting the acquisition offer is
\[
\pi_j^A(v_j) = \bar p_j^*(\bm v^*)-v_j+\int_0^{v_j}\overline{\mathcal{WP}}_j(x)dx.
\]

If seller $j$ rejects the acquisition offer, he participates in the procurement auction. Seller $j$ wins in the procurement auction if his valuation is lower than the reserve price and is 
 the lowest among all the sellers, and no seller sold to the speculator. Formally, seller $j$'s winning probability is 
\[
\mathcal{WP}_j(v_j):=\1_{\{v_j\leq r\}}\prod_{i\in\I\setminus\A}[1-F_i(v_j)]\prod_{s\in\A\setminus\{j\}}\left[1-F(\max\{v_j,v_s^*\})\right].
\]
Therefore, seller $j$'s expected payoff from the procurement auction is 
\[
\pi_j^R(v_j) = \int_{v_j}^r\mathcal{WP}_j(x)dx.
\]

Note that $\pi_j^A(v_j^*)=\pi_j^R(v_j^*)$. Then it suffices to show that $\pi_j^A(v_j)-\pi_j^R(v_j)$ decreases with $v_j$. In fact,
\[
\begin{split}
\frac{\partial[\pi_j^A(v_j)-\pi_j^R(v_j)]}{\partial v_j}   =-\bigg\{
&\prod_{s\in\A\setminus\{j\}}\left[1-F(\min\{v_j,v_s^*\})\right]\\
&-\1_{\{v_j\leq r\}}\prod_{i\in\I\setminus\A}[1-F_i(v_j)]\prod_{s\in\A\setminus\{j\}}\left[1-F(\max\{v_j,v_s^*\})\right]
\bigg\}
\leq0,
\end{split}
\]
where the inequality follows from $\1_{\{v_j\leq r\}}\prod_{i\in\I\setminus\A}[1-F_i(v_j)]\leq1$ and $1-F(\min\{v_j,v_s^*\})\geq1-F(\max\{v_j,v_s^*\})$. This completes the proof. 
\end{proof}

\begin{description}
\item[Proof of \Cref{no knock}]
\end{description}
\begin{proof}
Suppose that in an equilibrium, all the sellers with valuations lower than $r$ are knocked out of the procurement auction. Because $F(r)<1$, 
on the equilibrium path, rejection of the acquisition offer can still happen. Therefore, the speculator cannot tell whether a rejection is because of a deviation or because the seller has a valuation above $r$. So the speculator must bid $r$ in the procurement auction, regardless of how many rejections occurred. 

For a seller with value $v<r$, the equilibrium payoff in the compete knockout scenario is $\int_v^r[1-F(x)]^{N-1}dx$. However, if the seller deviates and bids $r-\epsilon$ in the procurement auction, where $\epsilon$ is an arbitrarily small number, he gets a higher payoff of $r-\epsilon-v$. This means complete knockout cannot happen in equilibrium.
\end{proof}

\end{appendices}

\clearpage
\bibliographystyle{aernoboldcomma}
\bibliography{speculation_ref}

\end{document}